\documentclass[graybox, vecphys]{article}

\usepackage{type1cm}        
\usepackage{makeidx}         
\usepackage{graphicx}        
\usepackage{multicol}        
\usepackage[bottom]{footmisc}

\usepackage{newtxtext}       %
\usepackage{newtxmath}       

\usepackage{booktabs}
\usepackage{multirow}
\usepackage{algorithm}
\usepackage{algpseudocode}
\usepackage{authblk}

\newcommand{\Q}{{\mathbb Q}}
\newcommand{\R}{{\mathbb R}}
\newcommand{\C}{{\mathbb C}}

\newcommand{\Z}{{\mathbb Z}}

\DeclareMathOperator*{\argmin}{argmin}

\DeclareMathOperator{\trace}{tr}
\DeclareMathOperator{\rank}{rk}

\DeclareMathOperator{\sgn}{sgn}

\renewcommand{\Re}{\operatorname{Re}}
\renewcommand{\Im}{\operatorname{Im}}

\makeindex             
\newtheorem{example}{Example}
\newtheorem{theorem}{Theorem}
\newtheorem{definition}{Definition}

\begin{document}
	
	\title{Recovery under Side Constraints}
\author[1]{Khaled Ardah\thanks{khaled.ardah@tu-ilmenau.de}}
\author[2]{Martin Haardt\thanks{martin.haardt@tu-ilmenau.de}}
\author[3]{Tianyi Liu\thanks{tliu@nt.tu-darmstadt.de}}
\author[4]{Frederic Matter\thanks{matter@mathematik.tu-darmstadt.de}}
\author[5]{Marius Pesavento\thanks{pesavento@nt.tu-darmstadt.de}}
\author[6]{ Marc E. Pfetsch\thanks{pfetsch@mathematik.tu-darmstadt.de}}
\affil[1,2]{Communications Research Laboratory, Ilmenau University of Technology, Helmholtzplatz 2, 98684 Ilmenau, Germany}
\affil[3,5]{Dept. of Electrical Engineering and Information Technology, TU Darmstadt, Merckstrasse 25, 64283 Darmstadt, Germany}
\affil[4,6]{Department of Mathematics, TU Darmstadt,
  Dolivostr. 15, 64293 Darmstadt, Germany}

%
%
\maketitle

\abstract{This chapter addresses sparse signal reconstruction under various types of structural side constraints with applications in multi-antenna systems. Side constraints may result from prior information on the measurement system and the sparse signal structure. They may involve the structure of the sensing matrix, the structure of the non-zero support values, the temporal structure of the sparse representation vector, and the nonlinear measurement structure. First, we demonstrate how a priori information in form of structural side constraints influence recovery guarantees (null space properties) using $\ell_1$-minimization. Furthermore, for constant modulus signals, signals with row-, block- and rank-sparsity, as well as non-circular signals, we illustrate how structural prior information can be used to devise efficient algorithms with improved recovery performance and reduced computational complexity. Finally, we address the measurement system design for linear and nonlinear measurements of sparse signals. Moreover, we discuss the linear mixing matrix design based on coherence minimization. Then we extend our focus to nonlinear measurement systems where we design parallel optimization algorithms to efficiently compute stationary points in the sparse phase retrieval problem with and without dictionary learning.}

\section{Introduction}
\label{HPPsec:Intro}
Compressed sensing (CS) is a signal processing technique for efficient acquisition and reconstruction of signals based on an underlying model sparsity, which allows to recover the signal of interest from far fewer samples than required by traditional acquisition systems operating at Nyquist rate. Theoretical recovery guarantees on the number of observations required can be further enhanced if side information on the measurement
system and the signal representation is incorporated in form of additional side constraints that are enforced in the recovery process. The measurement system may be subject to various types of side constraints that can be exploited and may originate from \emph{i)} the structure of the sensing matrix (shift-invariance, block structure, sparse co-array structures \cite{Steinwandt2017}, etc.), \emph{ii)} the structure of the sparse representation vector (integrality, variable bounds, unit-modulus, etc.),  \emph{iii)} the sparsity structure in the multiple snapshot case (block or group sparsity, rank sparsity, etc.), as well as  \emph{iv)} the structure of the measurements (quantization effects, K-bit measures, magnitude-only measurements, etc.). A fundamental question that arises in this context is, in which sense structural information can be incorporated into the CS problem and how it affects existing algorithms and theoretical results.

Moreover, recovery from nonlinear measurements with sparse models has recently been investigated, e.g., in the classical phase-retrieval problem, were different forms of redundancy have been incorporated through the use of known or unknown linear mixing networks. Redundancy can further enhance recovery in this case.

A large variety of applications involve data recorded from large-scale sensor arrays or \emph{massive multiple-input-multiple output} (MIMO) arrays, which consist of an assembly of wideband sensors to meet the corresponding high throughput and resolution requirements. In this context, sparsity naturally arises in the angular domain, e.g., in form of discrete propagation models and a small number of impinging waveforms from different directions. Similarly, in sensor array and MIMO applications, the structure of the array, the properties of the constellation signal and the transmitted waveform provide important prior information. In order to keep hardware costs in these large scale systems at a reasonable scale while retaining high performance, mixed analog-digital sensing system designs are employed to reduce the number and the sampling rates of the analog-to-digital converters as well as the quality requirements (e.g., w.r.t. linearity, dynamic range, etc.) of the hardware components.

This chapter reviews recent developments on sparse recovery guarantees and efficient recovery algorithms in CS networks under the aforementioned side constraints in the context of multi-antenna systems. First, CS with linear and nonlinear measurement models and the corresponding recovery problems are introduced in Section~\ref{sec:ProblemDescription}. Theoretical results on the recoverability of linear CS measurements under side constraints are presented in Section \ref{sec:RecoveryGuranteesUnderSideConstraints}. Recovery algorithms for sparse measurements under side constraints are addressed in Section \ref{sec:RecoveryAlgorithmsUnderSideConstraints}, and linear mixing matrix design is studied in Section \ref{sec:MixingMatrixDesign}. Finally, phase retrieval for known and unknown dictionaries is discussed in Section \ref{sec:RecoveryAlgorithmsForNonlinearMeasurements}, before conclusions are drawn in Section \ref{sec:ConclusionsAndOutlook}.

\section{Sparse Recovery in Sensor Arrays}\label{sec:ProblemDescription}

\begin{figure}
  \centering
  \includegraphics[scale=0.37]{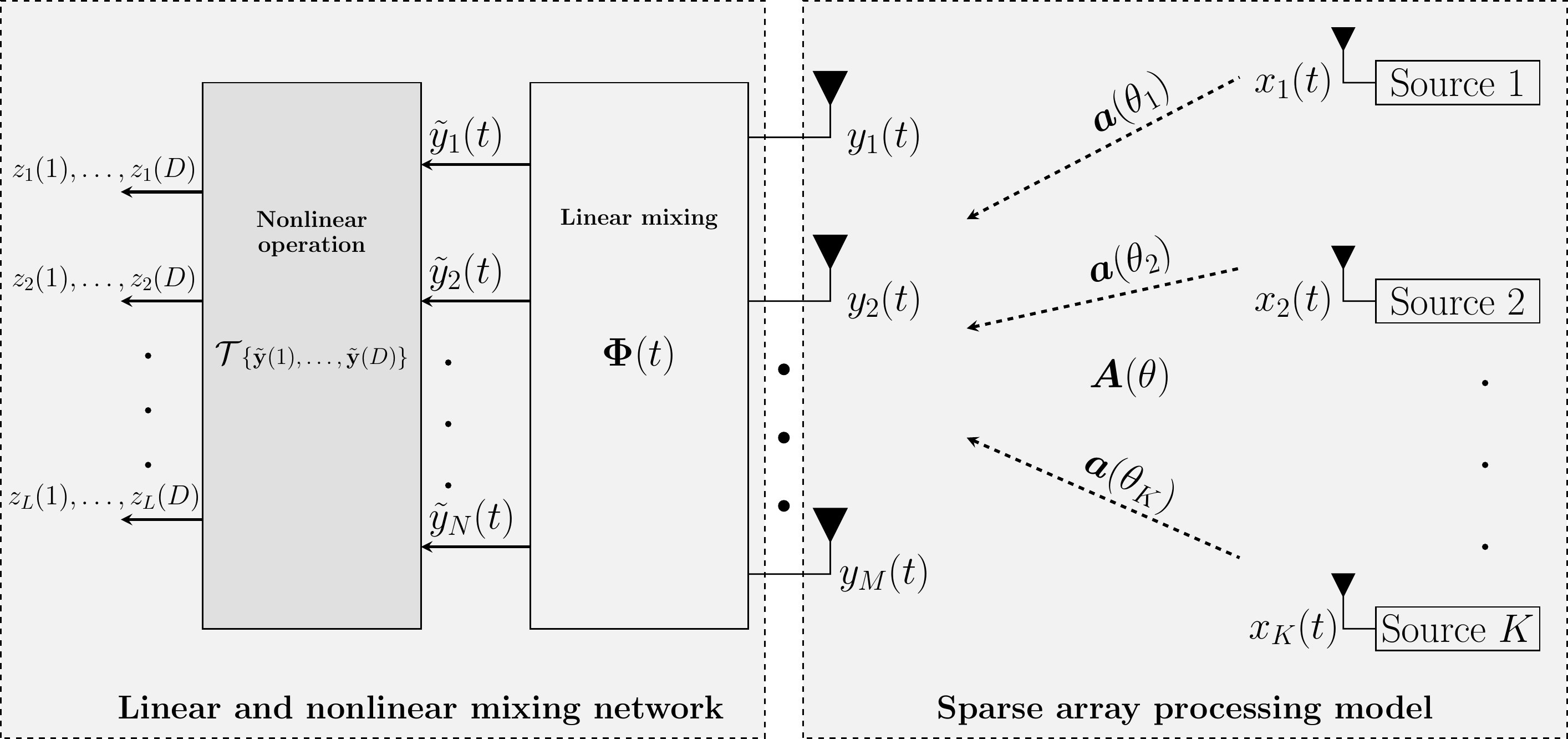}
  \caption{\small Sparse array processing model with linear and nonlinear mixing network.}
  \label{HPPfig:linear_nonlinear_mixing_network}
\end{figure}

Consider, as one prominent example application, the following sparse one-dim\-en\-sional narrow-band array processing model that is frequently encountered in the context of direction-of-arrival (DoA) estimation~\cite{KrimV96,
  VanTrees02, HaaPRK14,Ardah_spl2021,Ardah_wsa2019,Steinwandt2017gls} and multiple-input-multiple output (MIMO) communication~\cite{Gao19} and that will be used as a generic example in subsequent sections. We assume that $K$ far-field narrow-band source
signals impinge on a sensor array composed of $M$ omni-directional
sensors as depicted in the right-hand side of Fig.~\ref{HPPfig:linear_nonlinear_mixing_network}. The $t$-th time sample of the array output vector
${\bf y}(t) = [y_1(t), \ldots, y_M(t)]^{\rm T} \in \C^M$ is given by
\begin{equation} \label{HPPeq:matrix_model}
  {\bf y}(t) = {\bf A}({\boldsymbol \theta}^{(0)}) \, {\bf x}^{(0)}(t) +
  {\bf n}(t) , \qquad t = 1,\ldots, D,
\end{equation}
where
${\bf x}^{(0)}(t) = [x_1^{(0)}(t), \ldots, x_K^{(0)}(t)]^{\rm T} \in \C^K$
is the vector of waveforms emitted by the $K$ sources,
${\bf n}(t) \in \C^M$ contains the spatially and temporally white circular
Gaussian sensor noise, and $D$ is the number of available time samples. The
matrix
${\bf A}({\boldsymbol \theta}^{(0)}) = [{\bf a}(\theta_1),\ldots, {\bf
  a}(\theta_K)] \in \C^{M \times K}$ denotes the true array steering
matrix, whose $i$-th column is the array response vector
${\bf a}(\theta_i)$ corresponding to the $i$-th source with DoA
$\theta_i \in \Theta$, where $\Theta$ defines the field of view. The
steering vector ${\bf a}(\theta)$ describes a manifold denoted as
$\mathbb{M}^M$. For example, for a uniform linear array (ULA) with half-wavelength inter-element
spacing, ${\bf a}(\theta)$ is given by
${\bf a}(\theta)= [1,e^{-j \pi \sin (\theta)}, \ldots, e^{-j (M-1) \pi \sin
  (\theta)}]^{\rm T}$. We denote
${\boldsymbol \theta}^{(0)} = [\theta_1^{(0)}, \ldots, \theta_K^{(0)}]^{\rm
  T}$ as the true DoA parameter vector.

%
\subsection{Compressive Data Model for Sensor Arrays}
The model in~\eqref{HPPeq:matrix_model} presumes a dedicated radio frequency (RF)
receiver chain for each individual antenna element including an LNA,
filters, down-conversion, analog-to-digital converter (ADC), etc. In many
applications, however, such separate RF chains for each antenna element
come at a high cost in terms of the overall receiver complexity and
power consumption. To reduce the number of RF channels (and time samples)
without loss in the array aperture, compressed sensing can be applied,
where the antenna outputs are linearly combined in the analog domain and
then passed through a reduced number of RF chains to obtain the digital
baseband signals as illustrated in the left-hand side of
Fig.~\ref{HPPfig:linear_nonlinear_mixing_network}. This can be realized in
hardware, e.g., by using configurable hardware components such as tunable
phase shifters, a bank of fixed analog beamformers combined with a fast
switching network that enables analog beamformer selection, and/or a band
of (tunable) bandpass filters. This way, $N \leq M$ RF receiver channels are
used for signal processing in the digital domain.

Let $\boldsymbol{\Phi}^{(0)}(t) \in \C^{N \times M}$ denote the complex
analog mixing matrix of a compressive array at time $t$, which compresses
the output of $M$ antenna elements to $N$ active RF channels.  Then, the
complex (baseband) array output~\eqref{HPPeq:matrix_model} after combining can be
expressed as
\begin{equation} \label{HPPeq:compressed_matrix_model}
  \tilde{{\bf y}}(t) = \boldsymbol{\Phi}^{(0)}(t) \big( {\bf
      A}({\boldsymbol \theta}^{(0)}) \, {\bf x}^{(0)}(t) + {\bf n}(t)
  \big) + {\bf w}(t), \qquad t = 1,\ldots, D,
\end{equation}
where
$[\boldsymbol{\Phi}^{(0)} (t)]_{n,m} = \alpha_{n,m}(t) \cdot e^{j
  \varphi_{n,m}(t)},~n=1,\ldots,N,~m = 1,\ldots,M$ with
$\alpha_{n,m}(t) \in [0,1]$, $\varphi_{n,m}(t) \in [0,2\pi]$, and
${\bf w}(t) \in \C^N$ contains the spatially and temporally white circular
Gaussian measurement noise. Signals may be subject to additive noise that acts
before (i.e., in form of  ${\bf n}(t)$) or after the mixing network (i.e., in form of  ${\bf w}(t)$). Defining the effective array steering matrix
$\tilde{{\bf A}}({\boldsymbol \theta}^{(0)},t) = \boldsymbol{\Phi}^{(0)}(t)
{\bf A}({\boldsymbol \theta}^{(0)})$,
Model~\eqref{HPPeq:compressed_matrix_model} becomes
\begin{equation} \label{HPPeq:compressed_matrix_model2}
  \tilde{{\bf y}}(t) = \tilde{{\bf A}}({\boldsymbol \theta}^{(0)},t) \,
  {\bf x}^{(0)}(t) + \tilde{{\bf n}}(t),
\end{equation}
where
$\tilde{{\bf n}}(t) = \boldsymbol{\Phi}^{(0)}(t) {\bf n}(t) + {\bf w}(t)$
is the effective noise vector.

Cost efficient analog hardware devices and data acquisition systems
generally involve nonlinear transformations that can perform further
compression. Such nonlinear transformations are indicated by the operator
$\mathcal{T}$, which performs a nonlinear mapping from
$\mathbb{C}^{N \times D}$ to $\mathbb{C}^{L \times D}$ as depicted in
Fig.~\ref{HPPfig:linear_nonlinear_mixing_network}. The types of
nonlinearity consist, for instance, of nonlinear transformations introduced
from low-cost power amplifiers, magnitude-only and subband power
measurements that are often used in cellular communications, $C$-bit
quantization, the more aggressive 1-bit quantization (sign-only
measurements), hard-thresholding, and soft-thresholding, or modulo operations. Considering the
$D$ time samples simultaneously, the resulting measurement matrix
${\bf Z} = [ {\bf z}(1), \ldots, {\bf z}(D)] \in\C^{L \times D}$ recorded
at the output of the nonlinear mixing network is given by
\begin{equation} \label{HPPeq:compressed_matrix_model_nonlin}
  {\bf Z} = {\cal T} \big\{ \boldsymbol{\Phi}^{(0)}(1) {\bf
      A}({\boldsymbol \theta}^{(0)}) \, {\bf x}^{(0)}(1), \ldots,
    \boldsymbol{\Phi}^{(0)}(D) {\bf A}({\boldsymbol \theta}^{(0)}) \, {\bf
      x}^{(0)}(D) \big\} + {\bf N},
\end{equation}
where ${\bf N} \in\C^{L \times D}$ combines the various noise
contributions. If the mixing matrix $\boldsymbol{\Phi}^{(0)}(t)$ is
time-invariant, i.e.,
$\boldsymbol{\Phi}^{(0)}(t) = \boldsymbol{\Phi}^{(0)}$, the
model~\eqref{HPPeq:compressed_matrix_model_nonlin} reduces to
\begin{equation} \label{HPPeq:compressed_matrix_model_nonlin2}
  {\bf Z} = {\cal T} \big\{ \boldsymbol{\Phi}^{(0)} {\bf A}({\boldsymbol
      \theta}^{(0)}) \, {\bf X}^{(0)} \big\}  + {\bf N},
\end{equation}
where
${\bf X}^{(0)} = [{\bf x}^{(0)}(1), \ldots, {\bf x}^{(0)}(D)] \in\C^{K
  \times D}$ comprises the $D$ time snapshots.
\subsection{Sparse Recovery Formulations for Sensor Arrays}
Based on~\eqref{HPPeq:compressed_matrix_model_nonlin2}, we aim to solve the
sparse recovery problem that allows for a robust and efficient estimation
of the frequencies of the $K$ sources $\smash{x_k^{(0)}(t)}$ from the set of
measurements ${\bf z}(t)$ while exploiting potential structure in
$\boldsymbol{\Phi}(t)$, $\tilde{{\bf A}}$, ${\bf A}$, and
${\bf x}^{(0)}(t)$, or specific properties of~$\mathcal{T}$.  Specifically,
we will address variations of the general multiple measurement $\ell_{p,q}$ mixed-norm minimization problem
\begin{equation} \label{HPPeq:eq_sparse_recovery_problem} \tag{P0}
  \min_{{\bf X},\boldsymbol{\Phi}} \, \tfrac{1}{2} \left\| {\bf Z} -
    \mathcal T \big\{ \boldsymbol{\Phi} {\bf A}({\boldsymbol \theta}) {\bf
        X} \right\} \big\|_{\rm F}^2 + \lambda \lVert {\bf X}
  \rVert_{p,q} \,:\, \textit{side constraints},
\end{equation}
where at this point $\boldsymbol{\Phi}$ is assumed to be time-invariant for
simplicity of description
(i.e., considering~\eqref{HPPeq:compressed_matrix_model_nonlin2}),
${\bf A}({\boldsymbol \theta}) \in\C^{M \times P}$ with $P \gg M$ is a
``fat'' sensing matrix corresponding to the $P$-dimensional DoA grid vector
${\boldsymbol \theta}$ that appropriately samples the field of view~$\Theta$,
and ${\bf X} \in\C^{P \times D}$ is the row sparse (joint sparse)
signal matrix of interest, i.e., its columns share the same support.  The
support of the non-zero rows of ${\bf X}$ corresponds to the DoAs on the
spatial grid. Moreover, the regularization parameter $\lambda > 0$ controls
the trade-off between the data fitting term and the sparsity level in
${\bf X}$. The joint sparsity in ${\bf x}$ is induced by the $\ell_{p,q}$
mixed-norm defined as
\begin{align}
  \lVert {\bf X} \rVert_{p,q} = \Bigg( \sum_{k=1}^{P} \left\| {\bf x}_k
  \right\|_p^q \Bigg)^{1/q}, \label{HPPeq:lpqNorm}
\end{align}
for $p$, $q \geq1$, which applies an inner $\ell_p$ norm to the rows
${\bf x}_k$, $k=1,\ldots,P$ in
${\bf X} = [ {\bf x}_1, \ldots, {\bf x}_P]^{\rm T}$ and an outer $\ell_q$
norm to the $\ell_p$ row-norms. Ideally, we aim to
solve~\eqref{HPPeq:eq_sparse_recovery_problem} using the $\ell_{p,0}$ pseudo-norm
$\lVert {\bf X} \rVert_{p,0}$, which is the cardinality of the nonzero
$\ell_p$-norms of the rows of ${\bf X}$. If~$D = 1$, the model reduces to
the single measurement case and the $\ell_{p,1}$ mixed-norm reduces to the $\ell_1$ norm.

In the absence of the various noise contributions, i.e.,~${\bf N} = {\bf
  0}$, the general minimization problem~\eqref{HPPeq:eq_sparse_recovery_problem}
can be equivalently written as
\begin{equation}
  \label{HPPeq:P0_noiseless}
  \min_{{\bf X},\boldsymbol{\Phi}} \, \left\{\lVert {\bf X} \rVert_{p,q}
    \,:\, \mathcal T \left\{ \boldsymbol{\Phi} {\bf A}({\boldsymbol
        \theta}) {\bf X} \right\} = {\bf Z}, \textit{ additional
      side constraints}\right\}.
\end{equation}

\section{Recovery Guarantees under Side Constraints}\label{sec:RecoveryGuranteesUnderSideConstraints}

In this section, we consider the uniform recovery of sparse solutions with
additional side constraints on the solutions/signals. We use the signal model~\eqref{HPPeq:matrix_model} without noise in the single measurement case, i.e., $\mathbf{n} = {\bf 0}$ and $D=1$. More precisely,
consider the equation system $\mathbf{A x} = \mathbf{y}$ for
$\mathbf{A} \in \R^{m \times n}$, $\mathbf{y} \in \R^m$. The side
constraints for $\mathbf{x}$ can be expressed by requiring that
$\mathbf{x} \in C \subseteq \R^n$. This leads to optimization models
\begin{align}\label{prob:L_zero_w_sideconstraints}
  \min\, \{\lVert \mathbf{x} \rVert_0 \; : \; \mathbf{A x} = \mathbf{y}, \,
  \mathbf{x} \in C\},
\end{align}
i.e., variants of~\eqref{HPPeq:P0_noiseless} in the single measurement case without nonlinearities,
which promise to be able to uniquely recover sparse solutions for a larger
set of right hand side vectors~$\mathbf{y}$.  This is illustrated by the
following very simple toy example.

\begin{example}
  Consider the following recovery problem for $n = 2$. Let~$\mathbf{A} = [1, -1]$
  and~$y = 1$. The system $\mathbf{A} \mathbf{x} = y$ has two sparse solutions, namely~$\mathbf{x}_1 = (1,0)^{\rm T}$
  and~$\mathbf{x}_2 = (0,-1)^{\rm T}$.  Since~$\lVert \mathbf{x}_1 \rVert_1 = \lVert \mathbf{x}_2 \rVert_1 = 1$, it is not possible
  to \emph{uniquely} recover either point by~$\ell_1$-minimization, nor
  by~$\ell_0$-minimization. But by exploiting nonnegativity, $\mathbf{x}_1$ can
  indeed be uniquely recovered.
\end{example}

Another example of a whole family of sensing matrices showing that
exploiting side constraints leads to weaker recovery conditions can be found
in~\cite[Theorem~4.5]{HeuMPT20}. This shows that side constraints are not
only of theoretical interest, but should be exploited in the recovery
process. The price to pay may of course be that the recovery problems
become harder to solve.

\subsection{Integrality Constraints}

One particular example of an interesting side constraint is the integrality
of~${\bf x}$. Applications include discrete tomography~\cite{KuskeSP17} or massive MIMO with constellation
signals~\cite{HegdePP17,HegdeYSP16}.  A notable special case of this
setting includes the recovery of binary vectors, which has applications in
digital or wireless communication systems.

The corresponding general recovery problem can be formulated as
\begin{align}
  \label{eq:P0intbounds}
  \min\, \{\lVert \mathbf{x} \rVert_0 \, : \, \mathbf{A x} = \mathbf{A x}^{(0)}, \,
  \mathbf{x} \in [\boldsymbol{\ell}, \mathbf{u}]_\Z\},
\end{align}
where~$\mathbf{x}^{(0)} \in [\boldsymbol{\ell}, \mathbf{u}]_\Z \coloneqq
\{\mathbf{x} \in \Z^n \, : \, \ell_i \leq x_i \leq
u_i,\, i\in [n]\}$ is an~$s$-sparse vector and
$\mathbf{A} \in \R^{m \times n}$. Note that we can assume
$\boldsymbol{\ell} \in \Z^n \cup \{-\infty\}$ and
$\mathbf{u} \in \Z^n \cup \{\infty\}$. As in the case of classical sparse
recovery, we consider the~$\ell_1$-relaxation of~\eqref{eq:P0intbounds},
namely
\begin{align}
  \label{eq:P1intbounds}
  \min\, \{\lVert \mathbf{x} \rVert_1 \, : \, \mathbf{Ax} =
  \mathbf{Ax}^{(0)}, \, \mathbf{x} \in [\boldsymbol{\ell}, \mathbf{u}]_\Z\}.
\end{align}
In the literature, recovery of binary and integral sparse vectors
using~\eqref{eq:P1intbounds} has been considered for example
in~\cite{KeiperKLP17, Sto10}, where the nonconvex integrality condition was
relaxed to~$\mathbf{x} \in [\boldsymbol{\ell},\mathbf{u}] \coloneqq
\{\mathbf{x} \in \R^n \,:\, \boldsymbol{\ell} \leq \mathbf{x} \leq \mathbf{u}\}$. In this case,
the integrality assumption does not help for recovery: uniform recovery of
all sparse bounded integral~${\bf x}$ is equivalent to uniform recovery of
all sparse bounded~${\bf x}$, see~\cite{KeiperKLP17}. This already shows
that in order to exploit integrality, one has to take this into account in
the recovery program. Note that~\eqref{eq:P1intbounds} is nonconvex but can
be formulated as a mixed-integer (linear) program (MIP).

It turns out that in case of rational measurement matrices $\mathbf{A}$ and no bounds on
the variables, there is again no difference between integral and
general~${\bf x}$. However, in the presence of additional bounds, this is
no longer true. In this case it is possible to formulate null space properties
depending on the bounds~$\boldsymbol{\ell}$, $\mathbf{u}$ that characterize uniform recovery of
integral (bounded) sparse vectors~${\bf x}$ using~\eqref{eq:P1intbounds},
see~\cite{LangePST2016}. To this end, define the following two \emph{null
  space properties} (NSP) depending on a
set~$V \subseteq \R^n$. Let~${\bf A} \in \R^{m \times n}$ and~$S \subseteq
[n] \coloneq \{1,\dots,n\}$ and define
\begin{alignat*}{4}
  \text{NSP}(V):&& \qquad \lVert \mathbf{v}_S \rVert_1 < \lVert \mathbf{v}_{\overline{S}}
  \rVert_1 && \quad \forall \, \mathbf{v} \in (V \cap \mathcal{N}({\bf A})) \setminus
  \{\mathbf{0}\}, \\
  \text{NSP}_+(V):&& \qquad \mathbf{v}_{\overline{S}} \leq \mathbf{0} \; \implies \;
  \sum_{i=1}^n v_i < 0 &&  \quad \forall \, \mathbf{v} \in (V \cap
  \mathcal{N}({\bf A})) \setminus \{\mathbf{0}\},
\end{alignat*}
where~$\overline{S}$ denotes the complement of a set~$S$, $\mathbf{v}_S$ denotes the
vector of elements indexed by~$S$ and~$\mathcal{N}({\bf A})$ denotes the
null space of the matrix~${\bf A}$.

Then, NSP($\R^n$) is the classical null space property~\cite{GriN03,FouR13} which characterizes uniform recovery of
sparse vectors~${\bf x}$ by~$\ell_1$-minimization, and NSP$_+(\R^n)$ is the
well-known nonnegative null space property~\cite{KhaDXH11,Zha05}
characterizing uniform recovery via
nonnegative $\ell_1$-minimization.

For integral vectors without bounds, i.e., $\ell_i = -\infty$ and
$u_i = \infty$ for all $i \in [n]$, and integral nonnegative vectors, the
results for uniform recovery
are completely analogous to the classical case with the
only exception that for satisfying the NSP, only integral vectors in
the null space of~${\bf A}$ are of interest, see~\cite{LangePST2016} for
the exact statements. This observation also shows
that for~${\bf A} \in \Q^{m \times n}$, the classical (nonnegative) NSP and
the corresponding integral (nonnegative) NSP coincide. Thus, for rational
data, exploiting integrality does not lead to improved recovery
conditions.

If the bounds~$\boldsymbol{\ell}$, $\mathbf{u}$ are nontrivial, the situation changes
fundamentally. The first difference is that for classical recovery, bounds on~${\bf x}$ do not
influence recovery properties, since vectors in the null space
of~${\bf A}$ can be scaled accordingly. For integral vectors, however,
a new NSP in the presence of bounds~$-\infty \leq \ell_i
\leq 0 \leq u_i \leq \infty$ for all~$i\in [n]$ arises. Then it turns out that the
condition
NSP$([\boldsymbol{\ell}-\mathbf{u},\mathbf{u}-\boldsymbol{\ell}]_{\Z})$
provides a sufficient condition for uniform recovery
using~\eqref{eq:P1intbounds}, but not a characterization. Nevertheless we can use a variable split into
positive and negative part to obtain an NSP that characterizes uniform
recovery in the following statement.
\begin{theorem}[\cite{LangePST2016}]
  Let~${\bf A} \in \R^{m \times n}$ and~$s \geq 0$. Then every~$s$-sparse
  vector~$\mathbf{x}^{(0)} \in [\boldsymbol{\ell},\mathbf{u}]_\Z$ is the
  unique solution of~\eqref{eq:P1intbounds} if and only if
  \begin{align*}
    - (\mathbf{v}_{\overline{S}}, \mathbf{w}_{\overline{S}})^{\rm T} \in K \quad \implies \quad
    \sum_{i=1}^n v_i + w_i < 0,
  \end{align*}
  holds for all $(\mathbf{v},\mathbf{w})^{\rm T} \in \mathcal{N}({\bf A},-{\bf A}) \cap
  (K+(-K))$ with $(\mathbf{v},\mathbf{w})^{\rm T} \neq ({\bf 0},{\bf 0})^{\rm T}$
  and all~$S \subseteq [n]$, $\lvert S \rvert \leq s$, where
  \begin{align*}
    K \coloneqq \bigg\{ \binom{\mathbf{x}}{\mathbf{y}} \in
    \bigg[\binom{0}{0},\binom{\mathbf{u}}{-\boldsymbol{\ell}}\bigg]_{\Z}
    \, : \, x_i\cdot y_i = 0,\; i \in [n]\bigg\}.
  \end{align*} 
  \label{thm:NSP_integral_bounds}
\end{theorem}

The complementarity constraints~$x_i \cdot y_i = 0$ in $K$ are due to the
split into positive and negative part. This already shows that the
introduction of bounds leads to different recovery conditions, in contrast
to the situation of classical sparse recovery over~$\R^n$. For testing the
NSP in Theorem~\ref{thm:NSP_integral_bounds}, one needs to take
care of the complementarity constraints~$x_i \cdot y_i =0$. This can be
done by, e.g., using methods from~\cite{FisscherPfetsch18,FisP17a}.
For nonnegative integral vectors with upper bounds, the variable split is
not needed, and it can be shown that NSP$_+([-\mathbf{u},\mathbf{u}]_{\Z})$
characterizes uniform recovery~\cite{LangePST2016}.

Besides using~\eqref{eq:P1intbounds} for recovery of sparse integral
vectors, one can also use the exact recovery
problem~\eqref{eq:P0intbounds}, which can be formulated as a MIP if there are finite
bounds by expressing the nonconvex $\ell_0$-objective using
binary variables. In this case it also possible to characterize when
solving~\eqref{eq:P0intbounds} recovers any~$s$-sparse
integral vector with or without bounds. The condition of classical sparse recovery
is~$\text{spark}({\bf A}) > 2s$, where~$\text{spark}({\bf A})$
denotes the smallest number of linear dependent columns in~${\bf A}$. The
corresponding statements for integral sparse recovery appear
in~\cite{LangePST2016}.

\subsection{General Framework for Arbitrary Side Constraints}

In the previous section, we have explicitly considered integrality
constraints as one specific side constraint that can be exploited in the
recovery process. The corresponding recovery conditions resemble the
well-known null space properties that exist for various other settings such
as sparse (nonnegative) recovery~\cite{FouR13,GriN03,KhaDXH11,Zha05}, block-sparse recovery~\cite{StoPH09} or low-rank
(positive semidefinite) matrix recovery~\cite{KongSX14,OymH10}. Thus it seems reasonable to search
for a general setting and null space property that
unifies the cases already considered in the literature.
Such a general framework is presented in~\cite{JudKN14} that comprises all
the previously mentioned settings but
  does not handle additional side constraints such
as nonnegativity, integrality and positive semidefiniteness. Sparsity in
this general framework is expressed using projections. Recently, this
general framework was extended in~\cite{HeuMPT20} to also cover
additional side constraints. Under mild assumptions on the side constraints
and the measurement process it is possible to state an NSP for the
corresponding general recovery problem. It turns out that this general NSP
specializes to the already known NSPs in the various special cases
mentioned above. In the following, we will shortly describe this general
recovery framework and provide an application in order to evaluate the influence of side constraints.

For the general framework, we need two finite-dimensional Euclidean
spaces~$\mathcal{X}$ and $\mathcal{E}$. A \emph{linear sensing map}
$A \colon \mathcal{X} \to \R^m$ is used for acquiring signals
$x \in \mathcal{X}$ and a \emph{linear representation map}
$B \colon \mathcal{X} \to \mathcal{E}$ is used for mapping a signal to an
appropriate representation. We will denote the image of $x$ under a
linear operator $F$ as $Fx$. Additional side constraints are modeled using
a set~$\mathcal{C} \subseteq \mathcal{X}$ with~$0 \in \mathcal{C}$.
The image of~$\mathcal{C}$ under the map~$B$ is
denoted with~$\mathcal{D}$. Finally, let~$\lVert\cdot\rVert$ be a norm
on~$\mathcal{E}$.

Sparsity in this general framework is expressed using projections onto
appropriate subspaces. Therefore, let~$\mathcal{P}$ be a set of matrices
representing linear maps on $\mathcal{E}$. Each $P \in \mathcal{P}$ is assigned
a nonnegative real weight by $\nu \colon \mathcal{P} \to \R_+$ and a linear map
$\overline{P} \colon \mathcal{E} \to \mathcal{E}$.  Then, for~$s \in \R_+$,
an element~$\mathbf{x} \in \mathcal{X}$ is called~\emph{$s$-sparse}, if there exists
a linear map~$P \in \mathcal{P}$ with $\nu(P) \leq s$ and $PB\mathbf{x} =
B\mathbf{x}$. Furthermore, let
$\mathcal{P}_s = \{P \in \mathcal{P} \,:\, \nu(P) \leq s\}$ be the set
of linear maps that allow $s$-sparse elements.

The corresponding generalized recovery problem for a given right-hand
side~$\mathbf{y} \in \R^m$ can be formulated as
\begin{align}\label{eq:GeneralSettingRecoveryProblem}
  \min\, \{ \lVert B\mathbf{x}\rVert  \,:\,  A\mathbf{x} = \mathbf{y}, \, \mathbf{x} \in \mathcal{C} \}.
\end{align}
Note that this is convex if $\mathcal{C}$ is convex.
Using this general framework, it is possible to state two NSPs that can be
used to characterize uniform recovery using the general recovery
problem~\eqref{eq:GeneralSettingRecoveryProblem}.

\begin{definition}\label{def:GeneralSettingNSP}
  The linear sensing map $A$ satisfies the \emph{general null space
    property of type~I} and \emph{type II} of order $s$ for the set
  $\mathcal{C}$ if and only if for all
  $\mathbf{v} \in (\mathcal{N}(A) \cap (\mathcal{C}+(-\mathcal{C})))$ with
  $B\mathbf{v} \neq {\bf 0}$ and all $P \in \mathcal{P}_s$ it holds that
  \begin{equation}\label{eq:defGeneralSettingNSPa}
    \begin{aligned}
      -\overline{P}B\mathbf{v} \!\in\! \mathcal{D} \!\implies \!\exists \; \mathbf{v}^{(1)},
      \mathbf{v}^{(2)} \!\in\! \mathcal{C},\;
      \mathbf{v} = \mathbf{v}^{(1)} \!-\! \mathbf{v}^{(2)},\;
      \lVert PB\mathbf{v}^{(1)}\rVert - \lVert PB\mathbf{v}^{(2)}\rVert < \lVert
      \overline{P}B\mathbf{v} \rVert,
    \end{aligned}\tag{$\text{NSP-I}^\mathcal{C}$}
  \end{equation}
  \begin{equation}\label{eq:defGeneralSettingNSPb}
    \begin{aligned}
      -\overline{P}B\mathbf{v} \!\in\! \mathcal{D} \! \implies \! \forall \; \mathbf{v}^{(1)},\,
      \mathbf{v}^{(2)} \!\in\! \mathcal{C}, \mathbf{v} = \mathbf{v}^{(1)}\! - \!\mathbf{v}^{(2)},
      \lVert PB\mathbf{v}^{(1)}\rVert - \lVert PB\mathbf{v}^{(2)}\rVert < \lVert
      \overline{P}B\mathbf{v} \rVert,
    \end{aligned}\tag{$\text{NSP-II}^\mathcal{C}$}
  \end{equation}
  respectively,
  where~$\mathcal{N}(A) \coloneqq \{\mathbf{v} \in \mathcal{X} \,:\, A\mathbf{v} = {\bf 0} \}$ is
  the null space of~$A$.
\end{definition}

\begin{example}[{Recovery of sparse nonnegative vectors by $\ell_1$-minimization}] \ For the recovery of nonnegative vectors let\label{ex:LinearNonnegSettingDerivation}\
    $\mathcal{X} = \mathcal{E} = \R^n$, $B$ be the identity and
    $\lVert\cdot\rVert = \lVert\cdot\rVert_1$. The set of side constraints
    is $\mathcal{C} = \R^n_+$, implying $\mathcal{D} = \R^n_+$.  Let
    $\mathcal{P}$ be the set of orthogonal projectors onto all coordinate
    subspaces of $\R^n$, and define $\overline{P} \coloneqq I_n - P$, where
    $I_n$ denotes the identity mapping on $\R^n$.
    Define the nonnegative weight $\nu(P) \coloneqq \rank(P)$, so that
    $\nu(P)$ is the number of nonzero components of the subspace~$P$
    projects onto. The notion of general sparsity reduces to the classical
    sparsity of nonzero entries in a vector $\mathbf{x} \in \R_+^n$ and the recovery
    problem~\eqref{eq:GeneralSettingRecoveryProblem} becomes nonnegative
    $\ell_1$-minimization with $PB\mathbf{x} = \mathbf{x}_S$ and
    \smash{$\overline{P}B\mathbf{x} = \mathbf{x}_{\overline{S}}$}. In this case, it can be shown that
    the general null space property~\eqref{eq:defGeneralSettingNSPa} of
    order $s$ for the set $\mathcal{C}$ is equivalent to the known
    nonnegative null space property~\cite{KhaDXH11,Zha05}
    \begin{align}\label{eq:linearnonnegNSP}
      \hspace*{\leftmargin}
      \mathbf{v}_{\overline{S}} \leq 0 \, \implies \, \sum_{i \in S} v_i <
      \lVert \mathbf{v}_{\overline{S}}\rVert_1,  \; \forall \, \mathbf{v} \in \mathcal{N}(A)\setminus
      \{\mathbf{0}\}, \; \forall \, S \subseteq [n],\, \lvert S \rvert \leq s, \tag{$\text{NSP}_{\geq 0}$}
    \end{align}
    where $S$ denotes the index set of components on which $P$ projects.
\end{example}

Under mild assumptions, the null space
properties~\eqref{eq:defGeneralSettingNSPa}
and~\eqref{eq:defGeneralSettingNSPb} can be proven to characterize uniform
recovery using~\eqref{eq:GeneralSettingRecoveryProblem}. Which NSP is
needed depends on which assumptions are satisfied. For the formal statement
and more examples of how the various settings already considered in the
literature turn out to be special cases of this general recovery statement,
see~\cite{HeuMPT20}. At this point it is important to notice that already
in the special case of sparse vectors, checking whether~${\bf A}$ satisfies
the classical NSP is~$\mathcal{NP}$-hard~\cite{PfetschT14}.

The two NSPs characterizing uniform recovery in a very general framework
already indicate that a stronger, i.e., more restrictive side constraint
leads to weaker conditions that need to be satisfied to guarantee uniform
recovery.

In~\cite{HeuMPT20} an NSP for the
recovery of positive semidefinite block-diagonal matrices is derived, which has not
been considered before. Let~$\mathbf{X} \in \mathcal{S}^n_+$ be a
(symmetric) positive semidefinite matrix
and $\mathcal{A} \colon \mathcal{S}^n \to \R^m$, $\mathcal{A}(\mathbf{X}) =
(\mathbf{A}_1\bullet \mathbf{X}, \dots, \mathbf{A}_m \bullet \mathbf{X})^{\rm T}$ be a linear operator,
where $\mathbf{A}_1,\dots,\mathbf{A}_m \in \mathcal{S}^n$ are symmetric
matrices and ``$\bullet$'' denotes the componentwise inner product. In order to
define a block-diagonal form, let~$k \geq 1$
and~$B_1,\dots,B_k\neq \emptyset$ be a partition of~$[n]$. The matrix~$\mathbf{X}$
and the linear measurement operator~$\mathcal{A}(\mathbf{X})$ are in
\emph{block-diagonal form} with blocks~$B_1,\dots,B_k$,
if~$X_{s,t} = (A_i)_{s,t} = 0$ for
all~$(s,t) \notin (B_1 \times B_1) \cup \cdots \cup (B_k \times B_k)$ and
all~$i \in [m]$. Let~$\mathbf{X}_B$ be the the submatrix containing rows
and columns of~$\mathbf{X}$ indexed by~$B$. The corresponding norm is given by the $\ell_{*,q}$-norm
defined as
\begin{align*}
  \lVert \mathbf{X} \rVert_{*,q} \coloneqq \lVert (\lVert \mathbf{X}_{B_1}\rVert_*, \dots,
  \lVert \mathbf{X}_{B_k} \rVert_* )^{\rm T} \rVert_q,
\end{align*}
and the block-support $\text{BS}(X)$ is given by the indices of those
blocks~$\mathbf{X}_{B_i} \neq {\bf 0}$. By using an appropriate linear representation map
to encode the block-diagonal structure,~\eqref{eq:defGeneralSettingNSPa}
simplifies to
\begin{align}\label{eq:def_nsp_psd}
  \mathbf{V}_{B_i} \preceq 0 \; \forall \, i \in \overline{S} \quad \implies \quad
  \sum_{i \in S}  {\bf 1}^{\rm T} \lambda(\mathbf{V}_{B_i}) < \sum_{i \in \overline{S}}
  \lVert \mathbf{V}_{B_i} \rVert_{*}, \tag{$\text{NSP}_{*,1,\succeq 0}^*$}
\end{align}
for all $\mathbf{V} \in (\mathcal{N}(A) \cap \mathcal{S}^n) \setminus \{ \mathbf{0} \}$ and
all $S \subseteq [k]$, $\lvert S \rvert \leq s$, where $\lambda(\mathbf{V}_{B_i})$
is the vector of eigenvalues of $\mathbf{V}_{B_i}$, and~${\bf 1}$ is a vector of
ones. The general uniform recovery statement~\cite[Theorem 2.7]{HeuMPT20}
yields the following theorem.

\begin{theorem}[\cite{HeuMPT20}]
  Let $A(\mathbf{X})$ be a linear operator in block-diagonal form and $s \geq
  1$. Then, every positive semidefinite $\mathbf{X}^{(0)} \in \mathcal{S}^n_+$ with
  $ \lVert \mathbf{X}^{(0)} \rVert_{*,0} \leq s$ is the unique solution of
  $\min\, \{ \lVert \mathbf{X} \rVert_{*,1} \,:\, A(\mathbf{X}) = A(\mathbf{X}^{(0)}), \, \mathbf{X} \succeq 0
  \}$ if and only if $A(\mathbf{X})$ satisfies~\eqref{eq:def_nsp_psd} of order $s$.
\end{theorem}

As a conclusion, the general framework presented above can answer many interesting questions
concerning uniform recovery in the presence of side constraints using the
optimization problem~\eqref{eq:GeneralSettingRecoveryProblem}. The two
general null space properties~\eqref{eq:defGeneralSettingNSPa}
and~\eqref{eq:defGeneralSettingNSPb} can be used to analyze and quantify
the exact impact of various side constraints in the recovery process. Given
a specific setting, the NSPs can decide whether additional side information
is needed or which side constraints need to be exploited in the recovery
process to guarantee uniform recovery. For instance, this framework explains
why there are two seemingly different NSP formulations for
classical sparse recovery and nonnegative sparse recovery and their
connection.

\section{Recovery Algorithms Under Different Side Constraints for the
  Linear Measurement Model}\label{sec:RecoveryAlgorithmsUnderSideConstraints}

\subsection{Constant Modulus Constraints}
In this section we consider a variation of Problem~\eqref{prob:L_zero_w_sideconstraints} for the case of noisy measurements ${\bf s}$ and for side constraints on the sparse representation vector of the form  $\{\textbf{x} \in \C^N \,:\, \lvert x_n\rvert \in \{0,c\}\, \forall\, n \in [N]\}$. This problem emerges, e.g., in multi-user massive MIMO hybrid precoding systems with antenna selection and strict per antenna magnitude requirements~\cite{Fischer18}. In this application,
let~$\bf{A}$ denote the MIMO $N \times K$ channel matrix, ${\bf y}$ denote the symbol vector of the $K$ users, and $\textbf{x}$ denote the transmitted signal vector. To limit nonlinearity effects in the power amplifiers, the magnitudes of nonzero signals $x_n$ transmitted from the selected antennas are restricted to a constant $c$. The optimization problem can be formulated as~\cite{Fischer18}:
\begin{subequations}
	\begin{align}
		\min_{{\bf x} \in \mathbb{C}^N} \, &\lVert {\bf x} \rVert_0 \\
		\text{s.t.}\; &\lVert {\bf y} - {\bf A}^{\rm T}{\bf x}\rVert_2 \leq \sqrt{\delta}, \\
		&\,\lvert x_n \rvert \in \{0, c\}, \quad \forall n \in [N],\label{eq:AntennaSelectionCM}
	\end{align}\label{eq:AntennaSelection}
\end{subequations}
where $\lVert {\bf x}\rVert_0 = \lvert \{n\in [N] \,:\, x_n\neq 0\} \rvert$ denotes
the number of nonzero entries of~${\bf x}$, i.e., the number of active antennas. We assume without loss of generality that $c = 1$. 
In order to reformulate the constant modulus constraint \eqref{eq:AntennaSelectionCM}, we split vector ${\bf x}$ into real and imaginary part  $\Re \left[ {\bf x} \right]$ and $\Im \left[ {\bf x} \right]$, respectively. Let ${\bf b} = [b_1, b_2, \ldots, b_N]^{\rm T} \in \{0, 1\}^N$ denote a vector of binary variables. Problem~\eqref{eq:AntennaSelection} can then be written as
\begin{subequations}
	\begin{align}
          \min_{ {\bf x} \in \mathbb{C}^N, {\bf b} \in \{0,1\}^N } \,
          &\sum_{n=1}^N b_n \\
		\text{s.t.}\; &\sum_{k=1}^K \Big( \Re[y_k] -  \big(\Re[{\bf a}_k]^{\rm T}{\bf w}  - \Im[{\bf a}_k]^{\rm T}{\bf z}\big)    \Big)^2 \nonumber \\
		& \;\; +\Big( \Im[y_k] -  \big(\Re[{\bf a}_k]^{\rm T}\mathbf{z} + \Im[{\bf a}_k]^{\rm T}\mathbf{w} \big) \Big)^2 \leq {\delta},\quad \label{constraint:real_error_bound}\\
		& \Re \left[ x_n \right]^2 + \Im \left[ x_n \right]^2 \leq b_n, \quad \forall\, n \in [N], \label{eq:constraint:modulus_upper_bound}\\
		& \Re \left[ x_n \right]^2 + \Im \left[ x_n \right]^2 \geq b_n, \quad \forall\, n \in [N], \label{eq:constraint:modulus_lower_bound}\\
		& b_n \in \{0, 1\}, \,\qquad \forall\, n \in [N].
	\end{align}	\label{problem:real_domain_reformulation}
\end{subequations}
In \eqref{problem:real_domain_reformulation} we have replaced the modulus constraints $\lvert x_n\rvert^2 = \Re \left[ x_n \right]^2 + \Im \left[ x_n \right]^2 = b_n$, $n \in
[N]$, by the two inequality constraints~\eqref{eq:constraint:modulus_upper_bound}
and~\eqref{eq:constraint:modulus_lower_bound}, which will be treated differently in the following.
\begin{figure}[tb]
  \centering
  \includegraphics[width=0.8\textwidth]{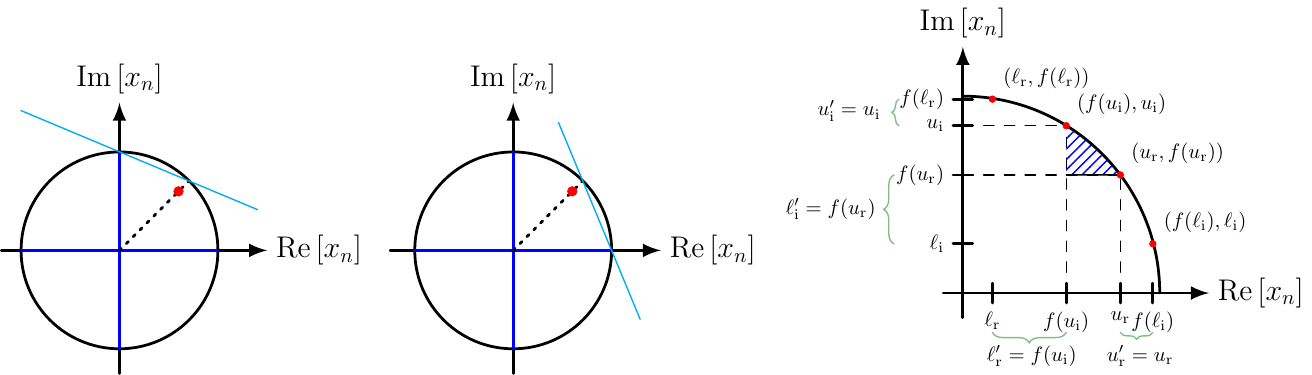}	
  \caption{\small Left: Inequalities that are added to the sub-nodes, Right: Bound propagation for the continuous variables.}
  \label{fig_propagation}
\end{figure}
The mixed-integer nonlinear program~\eqref{problem:real_domain_reformulation} will be solved by employing a spatial branching method~\cite{vigerske2013phd} in which branching is performed both on integral as well as continuous variables. 
In this branch-and-bound procedure the binary constraints $b_n \in \{0,1\} $
at each node of the tree are relaxed to linear inequality constraints $0 \leq b_n \leq 1$. 

In the case that the solution $(\hat{\bf x}, \hat{\bf b})$ of the LP relaxation of
Problem~\eqref{problem:real_domain_reformulation} does not satisfy the condition
$\Re[x_n]^2 + \Im \left[ x_n \right]^2 \geq b_n$ for some $n \in [N]$, this constraint violation will be resolved
by one of the following steps:
\begin{enumerate}
	\item If the binary variable $\hat{b}_n$ is already fixed to zero, the inequality $\Re[x_n]^2+\Im[x_n]^2 \leq b_n$ implies that also $\hat{x}_n$ is set to zero.
	\item If the bounds of the continuous variables $\Re[x_n]$ and $\Im[x_n]$ are not yet
	restricted to one of the orthants w.r.t.\ $\Re[x_n]\times \Im[x_n]$, four
	branching nodes can be created, the first with the additional constraints $\Re[x_n] \geq 0$,
	$\Im[x_n] \geq 0$, the second with $\Re[x_n]\geq 0$, $\Im[x_n] \leq 0$, the third with
	$\Re[x_n]\leq 0$, $\Im[x_n] \leq 0$, and the fourth with $\Re[x_n]\leq 0$, $\Im[x_n] \geq 0$.
	This partitions the feasible solution set into these four orthants.
	\item If the bounds of the continuous variables $\Re[x_n]$ and $\Im[x_n]$ are already
	restricted to one of these four orthants, the following steps are performed. Assume w.l.o.g.\ that
	$(\hat{x}_n , \hat{b}_n)$ is feasible for the first orthant,
	i.e., the one with $\Re[x_n] \geq 0$ and $\Im[x_n] \geq 0$.
	\item[] \textbf{Propagation:} Let $\ell_{\rm r} \leq \Re[ x_n ] \leq u_{\rm r}$,
		$\ell_{\rm i} \leq \Im[ x_n] \leq u_{\rm i}$ denote the current lower and upper bounds of
		the variables $\Re[x_n]$ and $\Im[x_n]$, respectively. Compute the four points $(\ell_{\rm r},f(\ell_{\rm r}))$,
		$(u_{\rm r}, f(u_{\rm r}))$, $(f(\ell_{\rm i}),\ell_{\rm i})$ and $(f(u_{\rm i}), u_{\rm i})$ on the unit circle
		that correspond to the respective lower and upper bounds of $\Re[x_n]$ and $\Im[x_n]$,
		where $f(x) = \sqrt{1 - x^2}$. These four points can now
		be used to strengthen the lower and upper bounds of~$\Re[x_n]$ and~$\Im[x_n]$. In
		order for an optimal solution $({\bf x}^{\star}, {\bf b}^{\star})$
		to fulfill the modulus constraint $\Re[x_n]^2 + \Im[x_n]^2 \geq b_n$, the point
		$(\Re[x_n^{\star}], \Im[x_n^{\star}])$ needs to lie on or above the arc between
		the two points ($\ell_{\rm r}',u_{\rm i}')$ and $(u_{\rm r}',\ell_{\rm i}')$ if $b_n^{\star} = 1$,
		where $\ell_{\rm r}' = \max \{\ell_{\rm r}, f(u_{\rm i})\}$, $
			u_{\rm r}' = \min \{u_{\rm r}, f(\ell_{\rm i})\}$, $\ell_{\rm i}' = \max \{\ell_{\rm i}, f(u_{\rm r})\}$, $u_{\rm i}' = \min \{u_2, f(\ell_{\rm r})\}$.
		This implies that the four values $\ell_{\rm r}'$, $u_{\rm r}'$, $\ell_{\rm i}'$ and $u_{\rm i}'$ can
		now be taken as new and possibly strengthened lower and upper bounds of
		$\Re[x_n]$ and $\Im[x_n]$, respectively.  If the binary variable $b_n$ is not yet
		fixed to one, the lower bounds are not propagated, as $b_n$ could be
		set to zero in an optimal solution, implying $\Re[x_n] = \Im[x_n] = 0$ as well.
		A visualization of this propagation is given in the right hand side of
		Fig.~\ref{fig_propagation}.
	\item[] \textbf{Separation:} If $\Re[\hat{x}_n] + \Im[\hat{x}_n] < \hat{b}_n$, add the
		cut $\Re[x_n] + \Im[x_n] \geq b_n$ to the LP relaxation. Note that each
		solution in this orthant on the unit circle satisfies this inequality.
              \item[] \textbf{Branching:} If $\Re[\hat{x}_n] + \Im[\hat{x}_n] \geq \hat{b}_n$, create
		two branching nodes defined by inequalities
		$f_n\, \Re[x_n] + g_n\, \Im[x_n] \geq b_n$, where $f_n \in \R$ and $g_n \in \R$
		can be computed according to the left hand side of
		Fig.~\ref{fig_propagation}.
\end{enumerate}
Computationally efficient suboptimal heuristic solutions for problem~\eqref{eq:AntennaSelection} and simulation results from numerical experiments can further be found in~\cite{Fischer18}.

\subsection{Row and Rank Sparsity}
In this section we consider row and rank sparse recovery from noisy measurements. 
\begin{figure}[tb]
  \centering
  \includegraphics[scale=1]{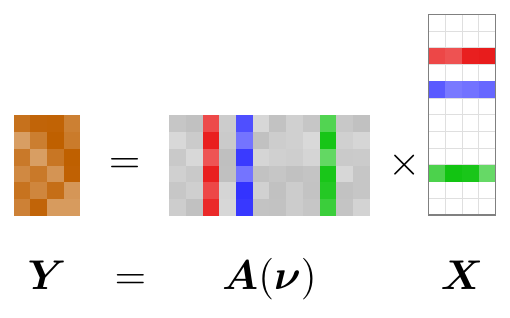}	
  \caption{\small Multiple measurement problem with row-sparsity.}
  \label{fig_rowsparse}
\end{figure}
The idea to exploit a common sparsity structure among multiple measurements as prior information was proposed in~\cite{10.2307/3647556, Tropp2006589, turlach2005simultaneous, Kowalski2009303, Malioutov:LassoDoa, 5466152}, where the mixed-norm~\eqref{HPPeq:lpqNorm} is used to enforce row-sparsity. The corresponding row-sparse data model is illustrated in Fig.~\ref{fig_rowsparse}. The classical row-sparse recovery problem corresponds to a least-squares
data fitting problem with $\ell_{2,1}$ mixed-norm minimization:
\begin{align}
	\min_{{\bf X}}\,
	\frac{1}{2} \left\| {\bf A} {\bf X} - {\bf Y} \right\|_{\rm F}^2 +
	\lambda\, \sqrt{D}\, \| {\bf X} \|_{2,1},
	\label{eq:mixedVectorNorm}
\end{align}
where ${\bf X} = [ {\bf x}(1),\ldots, {\bf x}(D) ]$. This problem emerges, e.g., in the context of Direction-of-Arrival (DoA) estimation, where the columns of the dictionary ${\bf A}$ represent the array responses for difference directions and the support of the matrix ${\bf X}$, i.e., the indices of the non-zero rows indicate the source DoAs.
The dimension of problem~\eqref{eq:mixedVectorNorm} grows with the number of measurements $D$ and the size of the dictionary and can become computationally intractable. To reduce the computational cost it was suggested in~\cite{Malioutov:LassoDoa} to reduce the dimension of the $M \times D$ measurement matrix ${\bf Y}$ by matching only the signal subspace in form of an $M \times K$ matrix ${\bf Y}_{\rm SV}$, leading to the prominent $\ell_1$-SVD method. A drawback of the $\ell_1$-SVD method is that it requires knowledge of the number of source signals and that the estimation performance may deteriorate in the case of correlated source signals. To overcome this limitation a convenient equivalent problem reformulation was derived in~\cite{Steffens_Sparrow_TSP} as stated in the following theorem.

\begin{theorem}[Problem Equivalence 1]
	The row-sparsity inducing $\ell_{2,1}$ mixed-norm minimization problem~\eqref{eq:mixedVectorNorm}
	is equivalent to the convex problem
	\begin{align}
		\min_{{\bf S} \in \mathbb{D}_+} &\,
		\trace \big( ({\bf A} {\bf S} {\bf A}^{\rm H}  + \lambda\, {\bf I}_M)^{-1} \hat{{\bf R}} \big) + \trace ({\bf S}), \label{eq:smr1}
	\end{align}
	with $\hat{{\bf R}} = {\bf Y} {\bf Y}^{\rm H} /D$ denoting the sample covariance matrix and $\mathbb{D}_+$ describing the set of nonnegative diagonal matrices, in the sense that minimizers ${\bf X}^\star$ and ${\bf S}^\star$ for problems~\eqref{eq:mixedVectorNorm} and~\eqref{eq:smr1}, respectively, are related by
	\begin{align}
		{\bf X}^\star = {\bf S}^\star \! {\bf A}^{\rm H}  ( {\bf A} {\bf S}^\star \! {\bf A}^{\rm H} + \lambda {\bf I}_M )^{-1} {\bf Y} .
		\label{eq:smr2}
	\end{align}
	Conversely, ${\bf S}^\star = {\rm diag}({s}_1^\star, \ldots, {s}_K^\star)$
	contains the row-norms of the sparse signal matrix ${\bf X}^\star = [{\bf x}_1^\star, \ldots, {\bf x}_K^\star]^{\rm T}$ on its diagonal according to
	\begin{align}
		{s}_k^\star= \frac{1}{\sqrt{D}} \| {\bf x}_l^\star \|_2,
		\label{eq:magIdentity}
	\end{align}
	for $k=1,\ldots,K$, such that the union support of ${\bf X}^\star$ is equivalently represented by the support of the sparse vector of row-norms $[{s}_1^\star, \ldots, {s}_K^\star]^{\rm T}$.
        \label{th:equivalence_}
      \end{theorem}
Problem~\eqref{eq:smr2} is known as the SPARse ROW-norm reconstruction
(SPARROW) reformulation. It reveals several interesting properties of the underlying multiple measurement problem and it can be reformulated as a semidefinite program. Unlike Problem~\eqref{eq:mixedVectorNorm}  the dimension of~\eqref{eq:smr2} does not grow with the number of measurements~\cite{Steffens_Sparrow_TSP}. Gridless variants of the method for uniform linear arrays (ULAs), shift-invariant arrays and augmentable arrays are reported in~\cite{Steffens_Sparrow_TSP,SuleimanAugmentable, SteffensGridlessSparrowShiftinvariant,Ardah_icassp2019,walewskiOffGridParameterEstimation2017}.
\begin{figure}[tb]
  \centering
  \includegraphics[scale=0.9]{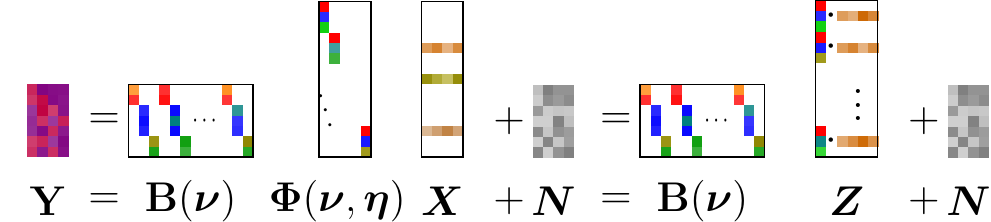}	
  \caption{\small Multiple measurement problem with block-sparsity.}
  \label{fig_blocksparse}
\end{figure}
In the case of DoA estimation in partly calibrated subarray systems with unknown DoAs~${\boldsymbol \nu}$ and subarray position parameters ${\boldsymbol \eta}$, the recovery problem can be formulated as a rank- and block-sparse regularization problem~\cite{SteffensCOBRAS}. The corresponding data model is illustrated in Fig.~\ref{fig_blocksparse}, where ${\bf B}({\boldsymbol \nu})$ contains the subarray steering vectors, ${\mathbf \Phi}({\boldsymbol \nu},{\boldsymbol \eta}) = [ {\boldsymbol \varphi}(\nu_1,{\boldsymbol \eta}),\ldots,  {\boldsymbol \varphi}(\nu_K,{\boldsymbol \eta})]$ contains the inter-subarray array responses, and ${\bf X}$ contains the row-sparse waveforms. We observe that the matrix
${\bf Z} = [{\bf Z}_1^{\rm T}, \ldots, {\bf Z}_K^{\rm T}]^{\rm T}$  enjoys a special block- and rank-sparse structure as it is composed of $K$ stacked rank-one matrices
${\bf Z}_k = {\boldsymbol \varphi}(\nu_k,{\boldsymbol \eta}) \, {\bf x}^{\rm T}_k$, for $k=1,\ldots,K$.
The block- and rank-sparse recovery problem is given by
\begin{equation}
	\min_{{\bf Z}} \,
	\frac{1}{2} \left\| {\bf B} {\bf Z} - {\bf Y} \right\|_{\rm F}^2 +
	\sum_{k=1}^K \| {\bf Z}_k \|_{*},
	\label{eq:NuclearNormProblem}
\end{equation} 
where the nuclear norm regularization $\| {\bf Z}_k \|_{*} = \trace \big( ({\bf Z}_k^{\rm H} {\bf Z}_k)^{1/2} \big)$ encourages block rank sparsity, i.e., the solutions blocks ${\bf Z}_k$ shall either be zero or low rank~\cite{yangParallelBestResponseAlgorithm2018,steffensMultidimensionalSparseRecovery2016,kusheParallelSparseRegularization2019,kusheBlockSuccessiveConvex2020}. Similar to Problem~\eqref{eq:mixedVectorNorm} also Problem~\eqref{eq:NuclearNormProblem} admits a convenient reformulation with a significantly reduced number optimization variables, as provided by the following theorem~\cite{SteffensCOBRAS,SteffensBookChapterSpectrum}.
\begin{theorem}[Problem Equivalence 2]
	The rank- and block-sparsity inducing $\ell_{*,1}$ mixed-norm minimization Problem~\eqref{eq:NuclearNormProblem} is equivalent to the convex problem
	\begin{align}
		\min_{{\bf S} \in {\cal S}_{K}^+}\, \trace \big( ({\bf B} {\bf S} {\bf B}^{\rm H}  + \lambda {\bf I})^{-1} \hat{{\bf R}} \big) + \trace ({\bf S}), \label{eq:smr1_}
	\end{align}
	with $\hat{{\bf R}} = {\bf Y} {\bf Y}^{\rm H} /D$ and ${\cal S}_{K}^+ $ denoting the sample covariance matrix and the set of positive semidefinite block-diagonal matrices composed of $K$ blocks of size $P \times P$, respectively. The equivalence holds in the sense that a minimizer  ${\bf Z}^\star$ for Problem~\eqref{eq:NuclearNormProblem} can be factorized as
	\begin{align}
		{\bf Z}^\star = {\bf S}^\star {\bf B}^{\rm H}  ( {\bf B} {\bf S}^\star {\bf B}^{\rm H} + \lambda {\bf I} )^{-1} {\bf Y},
		\label{eq:smr2_}
	\end{align}
	where ${\bf S}^\star$ is a minimizer for Problem~\eqref{eq:smr1_}.
        \label{th:equivalence2_}
\end{theorem}

\subsection{Block-Sparse Tensors}

In~\cite{BoHa:2016},  block-sparse core tensors were considered as the natural multidimensional
extension of block-sparse vectors in the context of multidimensional data acquisition.
The $( M_1, \dots , M_Q)$ block sparsity for a tensor assumes that $Q$ support sets,
characterized by $M_q$ indices corresponding to the non-zero entries, fully describe the
sparsity pattern of the considered tensor.
In the context of compressed sensing with Gaussian measurement matrices, 
the Cram\'{e}r-Rao bound (CRB) on the estimation accuracy of a Bernoulli-distributed
block-sparse core tensor was also derived in~\cite{BoHa:2016}.
This prior assumes that each entry of the core tensor has a given probability to
be non-zero, leading to random supports of truncated Binomial-distributed cardinalities.
Based on the limit form of the Poisson distribution, an approximated CRB expression
is provided for large dictionaries and a highly block-sparse core tensor. Using the
property that the $n$-mode unfoldings of a block-sparse tensor follow the
multiple-measurement vectors (MMV) model with a joint sparsity pattern, a fast and
accurate estimation scheme, called Beamformed mOde based Sparse Estimator (BOSE),
is proposed in the second part of~\cite{BoHa:2016}. The main contribution of
BOSE is to exploit the structure by mapping the MMV model onto the single-measurement vector (SMV) model, via beamforming
techniques. Finally, the proposed performance bounds and BOSE are applied in
the context of compressed sensing to non-bandlimited multidimensional signals with separable sampling
kernels and for multipath channels in a MIMO wireless communication scheme.

\subsection{Non-Circularity}

Recently, three different sparse recovery strategies have been proposed~\cite{SRH:16,SSPH:16,SRSHP:16} for exploiting the strict non-circularity property  of the impinging signals~\cite{Steinwandt2016nccrb,Steinwandt2017smoo}, i.e., the received complex symbols result from real-valued constellations 
rotated by an arbitrary phase $\phi$. As the rotation phase $\phi$ is usually unknown, the estimation 
problem becomes a two-dimensional (2-D) sparse recovery problem, which requires estimating the support 
in the spatial domain as well as in the rotation phase domain.

In~\cite{SRH:16}, a combined 2-D finite dictionary was introduced for both dimensions
and the resulting 2-D sparse recovery problem was solved by a $\ell_{2,1}$-mixed norm relaxation using multiple measurement vectors (MMV). Thereby,
the known benefits associated with strictly non-circular (NC) sources~\cite{Steinwandt2016nccrb,Steinwandt2017smoo}, e.g., an improved estimation accuracy and a doubling of the number of resolvable 
signals, can also be achieved via sparse recovery. In order to handle the resulting 2-D off-grid problem, an off-grid estimation procedure was introduced by means of local interpolation.

Article~\cite{SSPH:16} addresses the prohibitive computational complexity required for
solving the 2-D mixed-norm problem as a result of sampling both dimensions, significantly increasing
the size. Thus, in~\cite{SSPH:16} a sparse optimization framework was proposed based on nuclear
norm (rank) minimization after lifting the original optimization problem to a semidefinite
programming (SDP) problem in a higher-dimensional
space. To this end, the 2-D estimation problem is reduced to a 1-D estimation problem only in the sampled
spatial domain, which automatically provides grid-less estimates of the rotation phases. As a result, the
proposed method requires a significantly lower computational complexity while providing the same performance
benefits. Additionally, an off-grid estimator for the spatial domain has been proposed.

In~\cite{SRSHP:16}, a grid-less sparse recovery algorithm for NC signals has been proposed
based on atomic norm minimization (ANM). After the NC preprocessing step, the ANM-equivalent SDP problem
provides a solution matrix with a two-level Hermitian Toeplitz structure. It was shown 
that by using the multidimensional generalization of the Vandermonde decomposition, the desired direction
estimates can be uniquely extracted from the two-level Hermitian Toeplitz matrix via NC Standard ESPRIT or NC Unitary ESPRIT~\cite{HR_icassp:04}
in closed-form. 
Due to the exploitation of the NC signal structure, the proposed NC ANM procedure provides a superior estimation accuracy  as compared to the original methods for arbitrary signals. In this case, the number of estimated sources can exceed the number of sensors in the array.

\section{Mixing Matrix Design}\label{sec:MixingMatrixDesign}
In this section, we consider a noiseless time-invariant version of~\eqref{HPPeq:compressed_matrix_model} given as  
\begin{align}
	{\bf y} = {\bf \Phi} {\bf A} {\bf x} = {\bf \Psi} {\bf x} \in \C^{N} ,
\end{align}
where ${\bf \Psi} = {\bf \Phi} {\bf A} \in \C^{N\times P}$ is the total sensing matrix, ${\bf \Phi} \in \C^{N\times M}$ is the mixing matrix (a.k.a, the projection/compression matrix), ${\bf A} \in \C^{M\times P}$ is the dictionary matrix with $P \geq M$, and ${\bf x} \in \C^{P}$ is the signal vector of interest with $\Vert {\bf x} \Vert_{0} \leq s$,  i.e., ${\bf x}$ is $s$-sparse. To enhance recoverability of ${\bf x}$, the sensing matrix ${\bf \Psi}$ should be designed carefully so that it satisfies a certain property (e.g., the NSP or the RIP). Among them, the mutual coherence property of the sensing matrix ${\bf \Psi}$, denoted hereafter as $\mu_{\max}({{\bf \Psi}})$, provides an easy measure with respect to recoverability, which is defined as~\cite{CSTutorial}
\begin{align}\label{Mu}
	\mu_{\max}({\bf \Psi}) = \max_{i \neq j} \frac{|{\boldsymbol \psi}_i^{\rm H} {\boldsymbol \psi}_j|}{\Vert {\boldsymbol \psi}_i \Vert_2\, \Vert {\boldsymbol \psi}_j \Vert_2},
\end{align}
with columns ${\boldsymbol \psi}_k = [\psi_{k,1},\dots, \psi_{k,N} ]^{\rm T} \in \C^{N}$, $k \in \{1,\dots, P\}$. Clearly, a large coherence $\mu_{\max}({\bf \Psi})$ means that there exist, at least, two highly correlated columns in~${\bf \Psi}$, which may confuse any pursuit technique, such as BP and OMP. However, it has been shown that if $s < \frac{1}{2}\big(1 + {1}/{\mu_{\max}({\bf \Psi})} \big)$, the above techniques are guaranteed to recover ${\bf x}$ with high probability~\cite{CSTutorial,CSWCom}. Due to its simplicity, several sensing matrix design methods via mutual coherence minimization have been proposed recently, e.g., in~\cite{GDes,SVDShrinkage,ClosedForm}. In general, the results provided by~\cite{GDes,SVDShrinkage,ClosedForm} confirm that a well-designed sensing matrix always leads to a better recoverability. However, we note that the achievable mutual coherence by the aforementioned methods is, in general, far from the known {theoretical} Welch lower-bound, as we will also show in Section~\ref{Sec53}. Moreover, in the scenarios where the mixing matrix is realized using a network of phase shifters, none of the existing methods, to the best of our knowledge, have considered the constant modulus constraints imposed by the mixing matrix hardware that involves cost efficient analog phase shifters.

Formally, by assuming that the dictionary matrix ${\bf A} \in \C^{M\times P}$ is given and fixed, sensing matrix design reduces to finding the mixing matrix ${\bf \Phi}$ with constant modulus entries so that the coherence $\mu_{\max}({\bf \Psi})$ is minimized, which can be expressed as 
\begin{align}\label{MCM1}
	\underset{{\bf \Phi} \in \C^{N\times M}}{\min}\,  \mu_{\max}({\bf \Psi})\quad  \text{s.t.}\quad \Vert { {\boldsymbol \psi}_k } \Vert_2 = 1\; \forall k, \text{ and } \lvert\phi_{k,j}\rvert = 1\; \forall k,j.
\end{align}

Problem~\eqref{MCM1} is a non-convex and NP-hard optimization problem~\cite{Direct}. In the following, we propose two solution methods. Subsection~\ref{sec:smcm} presents the sequential mutual coherence minimization (SMCM) we proposed in~\cite{camsap2019} for the case of $P = M$. In Subsection~\ref{sec:general}, we propose a new method termed enhanced gradient-descent (EGD) for the more general case of $P \geq M$.

\subsection{Sensing Matrix Design: $P = M$ Case}\label{sec:smcm}

In this subsection, we present our first solution to problem~\eqref{MCM1} for unconstrained mixing matrix design, i.e., by neglecting the constant modulus constraints. Specifically, for a given dictionary matrix ${\bf A} \in \C^{M\times P}$, we assume that $P = M$ and the columns of ${\bf A}$ are linearly independent so that the condition of ${\bf A} {\bf A}^{-1} = {\bf I}_M$ is guaranteed. In this case, for a given sensing matrix ${\bf \Psi} \in \C^{N\times P}$ with a coherence $\mu_{\max} = \mu_{\max}({\bf \Psi})$, the optimal unconstrained mixing matrix that preserves $\mu_{\max}$ can be obtained as ${\bf \Phi}_\text{uncon} = {\bf \Psi} {\bf A}^{-1} \in \C^{N\times M}$, i.e., $\mu_{\max}({\bf \Phi}_\text{uncon} {\bf A}) = \mu_{\max}$. Therefore, the main task here is to find a low coherence sensing matrix ${\bf \Psi} \in \C^{N\times P}$.

Let us assume that the columns of ${\bf \Psi}$ are normalized so that $\Vert { {\boldsymbol \psi}_k } \Vert_2 = 1, \forall k$, and let ${\bf G} = {{\bf \Psi}}^{\rm H} {\bf \Psi}\in \C^{P\times P}$ be the so-called Gram-matrix of ${\bf \Psi}$. Moreover, let ${\bf G}_{\text{sqr-abs}} \in \R^{P\times P}$ be a matrix so that its $(k,j)$-th entry is given as ${\bf G}^{[k,j]}_{\text{sqr-abs}} = |{\bf G}^{[k,j]}|^2 $. By expanding ${\bf G}_{\text{sqr-abs}}$, it can be expressed as
\begin{align}\label{Gsquared}
	{\bf G}_{\text{sqr-abs}} = \begin{bmatrix}
		|{\boldsymbol \psi}_1^{\rm H} {\boldsymbol \psi}_1|^2  & \dots & |{\boldsymbol \psi}_1^{\rm H} {\boldsymbol \psi}_P|^2 \\
		\vdots & \ddots &  \vdots \\
		|{\boldsymbol \psi}_P^{\rm H} {\boldsymbol \psi}_1|^2  & \dots & |{\boldsymbol \psi}_P^{\rm H} {\boldsymbol \psi}_P|^2
	\end{bmatrix} = \begin{bmatrix}
		1  & \dots & |{\boldsymbol \psi}_1^{\rm H} {\boldsymbol \psi}_P|^2 \\
		\vdots & \ddots &  \vdots \\
		|{\boldsymbol \psi}_P^{\rm H} {\boldsymbol \psi}_1|^2  & \dots &1
	\end{bmatrix},
\end{align}
which is a symmetric matrix with all ones on its main diagonal. Since all vectors in ${\boldsymbol \Psi}$ have unit norm, we have ${\bf G}^{[k,j]}_{\text{sqr-abs}} = |{\boldsymbol \psi}_k^{\rm H} {\boldsymbol \psi}_j|^2 \leq 1, \forall k\neq j$, and the maximum among them represents the squared-coherence of the matrix ${\boldsymbol \Psi}$. According to~\cite{SteepestDest,Tropp}, $\mu_{\max}({\boldsymbol \Psi})$ has a theoretical lower bound given as $\mu_{\max}({\boldsymbol \Psi}) \geq \sqrt{\beta}$, where $\beta = {\frac{P-N}{N(P-1)}}$. This means that, at the best, we have $\mu_{\max}({\boldsymbol \Psi}) = \sqrt{\beta}$. Noting that the $k$-th column vector ${\boldsymbol \psi}_k$ appears only in the $k$-th column and row of ${\bf G}_{\text{sqr-abs}}$ (due to its symmetry), we propose to solve problem~\eqref{MCM1} in an alternating fashion by iterating over the following $P$ subproblems, where the $k$-th subproblem for updating ${\boldsymbol \psi}_k$ is given as   
\begin{equation}
	\begin{aligned}\label{MCM3}
		\text{find}\quad   {\boldsymbol \psi}_k \in \C^{N}\quad 
		\text{s.t.}\quad   |{\boldsymbol \psi}^{\rm H}_j {\boldsymbol \psi}_k|^2 \leq \beta\; \forall j \neq k, \text{ and }
		\Vert {\boldsymbol \psi}_k  \Vert_2 = 1.
	\end{aligned}
\end{equation}

Problems~\eqref{MCM1} and~\eqref{MCM3} are related in the sense that both aim to minimize the maximum off-diagonal entry in~\eqref{Gsquared}. However, the strict unit-norm constraint $\Vert  {\boldsymbol \psi}_k \Vert_2 = 1$ in Problem~\eqref{MCM3} may result in infeasibility for poorly initialized vectors~${\boldsymbol \psi}_j,\forall j\neq k$, especially with a tight lower-bound $\beta$. To avoid such a scenario, we propose to relax~\eqref{MCM3} by dropping the unit-norm constraint and only impose it after a solution is obtained, i.e., we first seek a solution to the following relaxed problem 
\begin{equation}
	\begin{aligned}\label{MCM4}
		\text{find}\quad   {\boldsymbol \psi}_k \in \C^{N} \quad  \text{s.t.}\quad  | {\boldsymbol \psi}^{\rm H}_j  {\boldsymbol \psi}_k|^2 \leq \beta\; \forall j \neq k,
	\end{aligned}
\end{equation}
which, unlike~\eqref{MCM3}, is guaranteed to be feasible. To obtain a solution of problem~\eqref{MCM4}, a suitable objective function is needed. One possible approach is as follows 
\begin{equation}
	\begin{aligned}\label{MCM8}
		{\boldsymbol \psi}_k \in \underset{{{\bf v}_k \in \C^{N}}}{\arg \max} \;  |{\boldsymbol \psi}_k^{\rm H} {\bf v}_k|^2 \quad
		\text{s.t.}\quad  |{\boldsymbol \psi}^{\rm H}_j {\bf v}_k|^2 \leq \beta\; \forall j \neq k,
	\end{aligned}
\end{equation}

\begin{figure}[tb]
	\centering
	\includegraphics[width=0.5\linewidth]{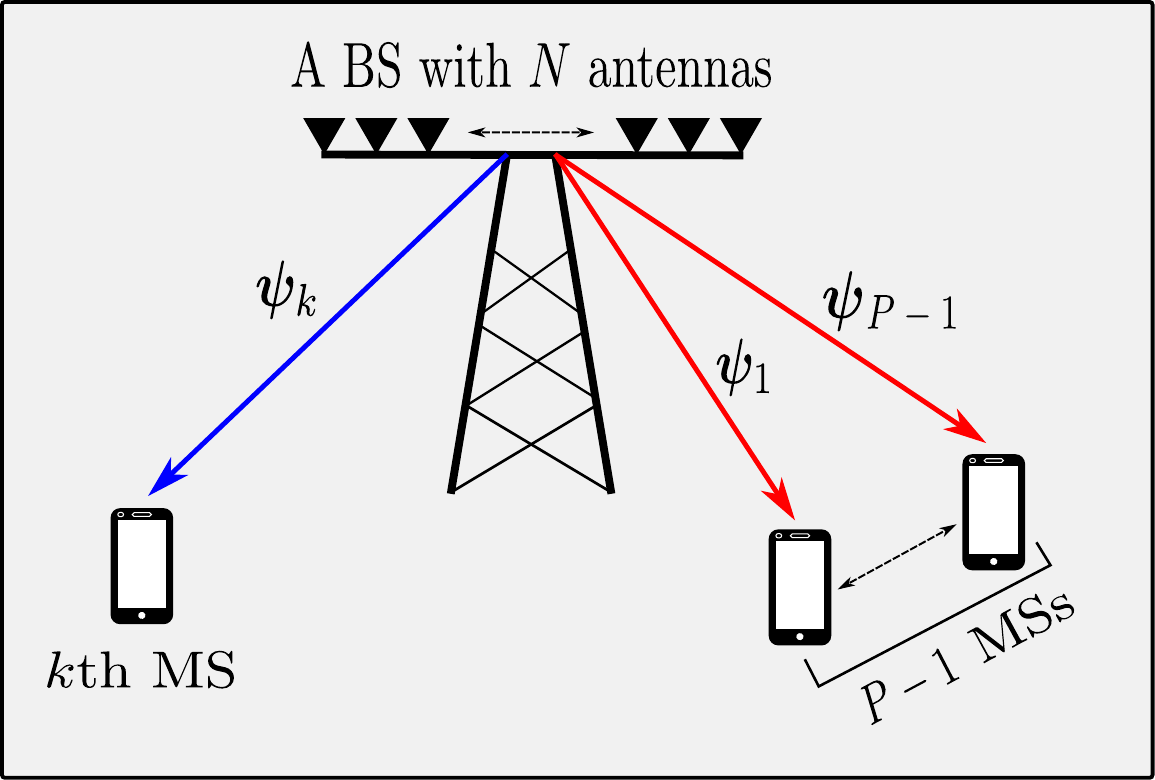}
	\caption{\small A $P$-user interference-channel (IC) system model in wireless communication systems, where a base-station (BS) with $N$ antennas serves $P$ single-antenna mobile-stations (MSs) simultaneously so that the data transmission to the $k$-th MS causes interference to the remaining $P-1$ MSs.}
	\label{fig:kic}
\end{figure}

In problem~\eqref{MCM8}, we borrow the notion from the beamforming design in wireless communication systems, see Fig.~\ref{fig:kic}, where we interpret ${\bf v}_k \in \C^{N}$ as the beamforming vector of the $k$-th mobile station (MS) that we wish to design so that the desired transmit signal to the $k$-th MS, i.e., $|{\boldsymbol \psi}_k^{\rm H} {\bf v}_k|^2$, is maximized and the interference signals to the remaining $P-1$ MSs, i.e., $|{\boldsymbol \psi}^{\rm H}_j {\bf v}_k|^2 \leq \beta, \forall j \neq k$, are minimized for given channel vectors $\{{\boldsymbol \psi}_1, \dots,{\boldsymbol \psi}_P\}$. Due to its convexity, Problem~\eqref{MCM8} can be efficiently solved using existing techniques, e.g., using the proposed method in~\cite{SOP}, as we have shown in~\cite{camsap2019,Ardah_icassp2020}. Alternatively, we can resort to the relaxed semidefinite programming (SDP) approach, by dropping the rank-one constraint, and write Problem~\eqref{MCM8} as 
\begin{equation}
	\begin{aligned}\label{MC_SDP}
		\max_{{\bf V}_k \in \C^{N \times N }}\;  \trace\{ {\boldsymbol \Psi}^{\text{cov}}_k {\bf V}_k  \} \quad
		\text{s.t.}\quad  \trace \{ {\boldsymbol \Psi}^{\text{cov}}_j {\bf V}_k  \} \leq \beta\; \forall j \neq k,
		\text{ and } {\bf V}_k  \succeq 0,  
	\end{aligned}
\end{equation}
where ${\boldsymbol \Psi}^{\text{cov}}_k = {\boldsymbol \psi}_k {\boldsymbol \psi}_k^{\rm H} \in \C^{N \times N }$ and ${\bf V}_k = {\bf v}_k {\bf v }^{\rm H}_k \in \C^{N \times N }$. Problem~\eqref{MC_SDP} is convex and can be efficiently solved using off-the-shelf solvers, e.g., the CVX toolbox. Let ${\bf V}_k$ denote the obtained solution of~\eqref{MC_SDP}. Then, ${\boldsymbol \psi}_k$ is given by the eigenvector corresponding to the dominant eigenvalue of ${\bf V}_k$, i.e., ${\boldsymbol \psi}_k =  \lambda_{\text{max}} \{ {\bf V}_k  \}$. In summary, the proposed mixing matrix design method is given by Algorithm~\ref{SeqMCM2}. Note that a na\"ive approach to obtain a constrained mixing matrix, i.e., one with constant modulus entries, is given as ${\bf \Phi}_\text{con} = {\bf \Pi}({\boldsymbol \Phi}_\text{uncon})$, where ${\bf \Pi}(\cdot)$ is a projection function that imposes the constant modulus constraints on ${\boldsymbol \Phi}_\text{uncon}$ element-wise, i.e., ${\bf \Pi}(z) = z/|z|$. The performance of such an approach will also be evaluated in Section~\ref{Sec53}.   

\begin{algorithm}[t]
	\caption{Sequential mutual coherence minimization (\textsf{SMCM}) }
	\label{SeqMCM2}
	{\footnotesize 
		\begin{algorithmic}[1]
			\State{\textbf{Inputs}: ${\boldsymbol \Psi}_{(0)} \in \C^{N\times P}$. Select $\epsilon_{\text{th}}$. Set  $\beta = {\frac{P-N}{N(P-1)}}$ and $n = 1$.}
			\For{$n = 1, 2, \dots$}
			
			\For{$k = 1$ to $P$}	
			\State{1: Compute ${\bf V}_{k(n)}$ by solving problem~\eqref{MC_SDP}.}
			
			\State{2: Update the $k$-th column vector of ${\boldsymbol \Psi}_{(n)}$ as ${\boldsymbol \psi}_{k(t)} =  \lambda_{\text{max}} \{ {\bf V}_{k(n)} \}$.}
			
			\EndFor
			
			\If{$\epsilon =  |\mu_{\max}({\boldsymbol \Psi}_{(n)}) - \mu_{\max}({\boldsymbol \Psi}_{(n-1)})|^{2} \leq \epsilon_{\text{th}}$}
			\State{Break}
			\EndIf
			
			\EndFor
			\State{\textbf{Output:} the sensing matrix ${\boldsymbol \Psi}$ and the corresponding mixing matrix ${\boldsymbol \Phi}_\text{uncon} = {\boldsymbol \Psi} {\bf A}^{-1}$}. \label{step13}
		\end{algorithmic}
	}
\end{algorithm}

\subsection{Sensing Matrix Design: The General Case}\label{sec:general}

In this subsection, we propose a new solution to~\eqref{MCM1} for the more general case of $P \geq M$. Similarly to~\cite{GDes}, we propose to solve~\eqref{MCM1} indirectly by solving 
\begin{align}\label{MCM_F}
	\underset{{\bf \Phi} \in \C^{N\times M}}{\min} \;    \eta( {\bf \Phi}  ) \quad \text{s.t.}\quad \Vert { {\boldsymbol \psi}_k } \Vert_2 = 1\; \forall k, \text{ and } |\phi_{k,j}| = 1\; \forall k,j,
\end{align} 
where $\eta( {\bf \Phi}  ) = \Vert  {\bf A}^{\rm H} {\bf \Phi}^{\rm H} {\bf \Phi} {\bf A} - {\bf I}_{P}  \Vert^2_{\rm F}$. To obtain a solution for~\eqref{MCM_F}, we propose a constrained gradient-descent (GD) method, which updates the mixing matrix ${\bf \Phi}$ iteratively as 
\begin{align}\label{phit}
	{\bf \Phi}_{(n)} = {\bf \Pi} \bigg( {\bf \Phi}_{(n-1)} - \zeta \cdot  \frac{\partial \eta( {\bf \Phi}_{(n-1)}  )}{\partial {\bf \Phi}_{(n-1)}}\bigg),
\end{align}
where $n$ is the iteration index, $\zeta$ is the step-size, and $\frac{\partial \eta( {\bf \Phi}_{(n-1)}  )}{\partial {\bf \Phi}_{(n-1)}}$ is the gradient of $\eta( {\bf \Phi}_{(n-1)}  )$ with respect to ${\bf \Phi}_{(n-1)}$, which is given as~\cite{GDes}
\begin{align}\label{GradPhi}
	\frac{\partial \eta( {\bf \Phi}_{(n-1)}  )}{\partial {\bf \Phi}_{(n-1)} } =  {\bf \Phi}_{(n-1)} {\bf A} ( {\bf A}^{\rm H} {\bf \Phi}^{\rm H}_{(n-1)} {\bf \Phi}_{(n-1)} {\bf A} - {\bf I}_{P}  ) {\bf A}^{\rm H} = {\bf \Psi}_{(n-1)} {\bf E}_{(n-1)} {\bf A}^{\rm H},  
\end{align}
where ${\bf \Psi}_{(n-1)} = {\bf \Phi}_{(n-1)} {\bf A}$ and ${\bf E}_{(n-1)} = {\bf \Psi}^{\rm H}_{(n-1)} {\bf \Psi}_{(n-1)} - {\bf I}_{P} $. The update step in~\eqref{phit} is a direct extension of the proposed unconstrained GD method in~\cite{GDes} to account for the constant modulus constraints. Our results show that both the unconstrained and the constrained GD-based methods achieve a mutual coherence that is far from the known theoretical Welch lower-bound, as it is shown in Table~\ref{MC_Tab}. To enhance their performance, we propose to apply a shrinking operator on the error matrix ${\bf E}_{(n-1)}$ entry-wise to get $\tilde{{\bf E}}_{(n-1)}$ such that the $(k,j)$-th entry of $\tilde{{\bf E}}_{(n-1)}$ is obtained as  		
\begin{align}\label{ShrinkE}
	\tilde{{\bf E}}^{[k,j]}_{(n-1)}  = \begin{cases}
		0, & \big|	{\bf E}^{[k,j]}_{(n-1)} \big| < \alpha\cdot\sqrt{\beta},  \\
		\sgn \{ 	{\bf E}^{[k,j]}_{(n-1)}   \} \cdot \big( \big|	{\bf E}^{[k,j]}_{(n-1)} \big| - \alpha \cdot \sqrt{\beta} \big), & \text{otherwise},
	\end{cases}
\end{align}
where $\alpha \geq 1$ is an uncertainty measure and $\beta$ is as defined above. After a closer look at~\eqref{ShrinkE}, one can see that for a very tight threshold $ \bar{\beta} = \alpha \cdot\sqrt{\beta}$, the resulting error matrix $\tilde{{\bf E}}_{(n-1)}$ becomes a sparse matrix, where some of its entries that are smaller than $\bar{\beta}$ will be set to zero. The direct implication of such a shrinking operator is that the new mixing matrix ${\bf \Phi}_{(n)}$ will be updated so that it mainly minimizes the entries that are larger than $\bar{\beta}$. In summary, the proposed enhanced GD (EGD) method for mixing matrix design is given by Algorithm~\ref{EnhancedGrad}. In Section~\ref{Sec53}, we will investigate in detail the impact of $\alpha$ on the performance of EGD method.   

\begin{algorithm}[t]
	\caption{Enhanced gradient-descent (EGD)}
	\label{EnhancedGrad}
	{\footnotesize 
		\begin{algorithmic}[1]
			\State{\textbf{Inputs}: ${\boldsymbol \Phi}_{(0)} \in \C^{N\times M} $ and ${\bf A} \in \C^{M\times P} $. Select $\epsilon_{\text{th}} $ and $ \zeta$. Set $\beta = {\frac{P-N}{N(P-1)}}$ and $n = 1$.}
			\State{Normalize the columns of  ${\bf \Psi}_{(0)} = {\boldsymbol \Phi}_{(0)} {\bf A}$ so that $\Vert {\boldsymbol \psi_{(0),k} } \Vert_{2} = 1, \forall k$.}
			\For{$n = 1, 2, \dots$}
			\State{Calculate the error matrix ${\bf E}_{(n-1)} = {\bf \Psi}^{\rm H}_{(n-1)} {\bf \Psi}_{(n-1)} - {\bf I}_{P} $.}
			
			\State{Apply the shrinking operator~\eqref{ShrinkE} on ${\bf E}_{(n-1)} $ to get $\tilde{{\bf E}}_{(n-1)}$.}

			\If{mixing matrix should be unconstrained (i.e., ${\boldsymbol \Phi}_\text{uncon}$)}
			\State{Compute ${\bf \Phi}_{(n)} = {\bf \Phi}_{(n-1)} - \zeta \cdot {\bf \Psi}_{(n-1)} \tilde{{\bf E}}_{(n-1)} {\bf A}^{\rm H}  $.}
			\ElsIf{mixing matrix should be constrained (i.e., ${\boldsymbol \Phi}_\text{con}$)}
			\State{Compute ${\bf \Phi}_{(n)} = {\bf \Pi} \big( {\bf \Phi}_{(n-1)} - \zeta \cdot {\bf \Psi}_{(n-1)} \tilde{{\bf E}}_{(n-1)} {\bf A}^{\rm H}  \big)$.}
			\EndIf
			
			\State{Normalize the columns of  ${\bf \Psi}_{(n)} = {\boldsymbol \Phi}_{(n)} {\bf A}$ so that $\Vert {\boldsymbol \psi_{(n),k} } \Vert_{2} = 1, \forall k$.}

			\If{$\epsilon =  |\mu({\bf \Psi}_{(n)}) - \mu({\bf \Psi}_{(n-1)})|^{2} \leq \epsilon_{\text{th}}$}
			\State{Break}
			\EndIf
			\EndFor
			\State{\textbf{Output:} Mixing matrix ${\boldsymbol \Phi}^{\star}$}
		\end{algorithmic}
	}
\end{algorithm}

\subsection{Numerical Results}\label{Sec53}
\begin{figure}[tb]
	\centering
	\includegraphics[width=0.45\linewidth]{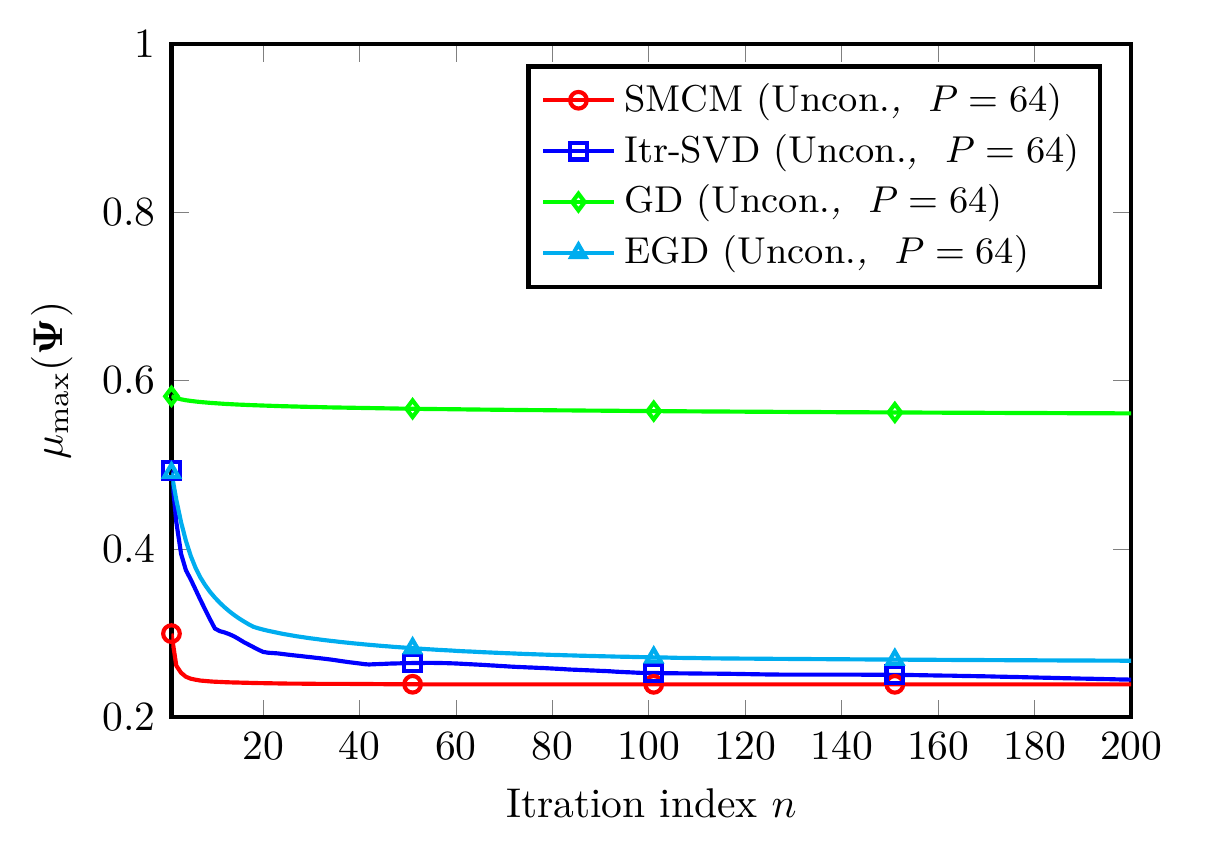}%
	\includegraphics[width=0.45\linewidth]{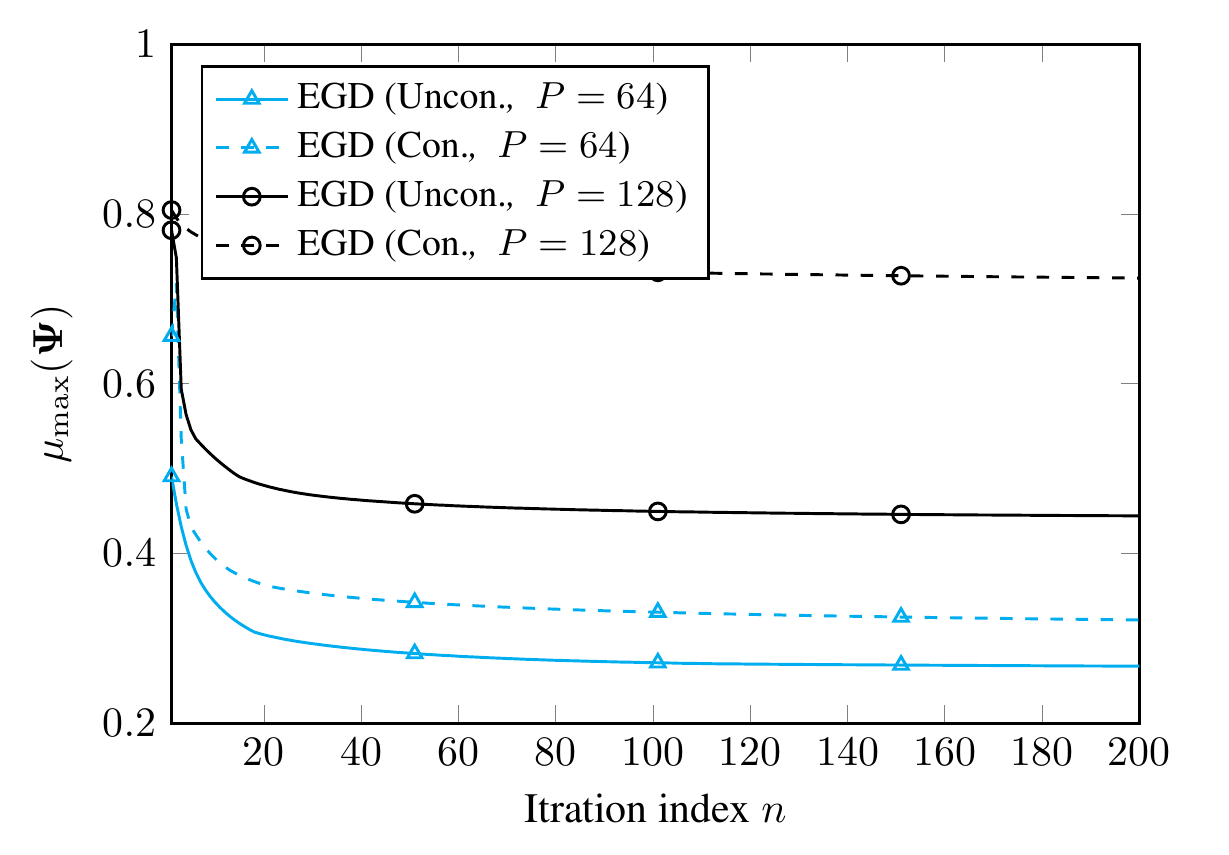}
	\caption{\small Coherence $\mu_{\max}{({\bf \Psi})}$ versus the iteration index.}
	\label{conv1}
\end{figure}

In this subsection, we present some numerical results for the proposed sensing matrix design methods. In all the simulation results, we set $N = 16$, $M = 64$, and design the dictionary matrix as ${\bf A} = [{\bf a}_1 , \dots, {\bf a}_P] \in \C^{M\times P}$ such that its $k$-th column is given as ${\bf a}_k = [1, e^{j \nu_k},\dots, e^{j \nu_k(M-1)} ]^{\rm T} \in \C^{M}$, where $\nu_k = (2\pi (k-1))/P$. For comparison, we include results for a mixing matrix ${\boldsymbol \Phi}$ obtained by using the proposed closed-form method in~\cite{ClosedForm}\footnote{Let ${\bf U} {\boldsymbol \Lambda} {\bf U}^{\rm H}$ be the eigenvalue decomposition of $ {\bf A}^{\rm H} {\bf A}$. Then, the unconstrained mixing matrix is obtained as \smash{${\bf \Phi}_{\text{uncon}} = {\boldsymbol \Lambda}^{-1/2}_N {\bf U}^{\rm H}_N$}, where ${\boldsymbol \Lambda}_N$ and ${\bf U}_N$ contain the leading $N$ eigenvalues and eigenvectors, respectively. For constrained mixing matrix scenarios, simply ${\bf \Phi}_\text{con} = {\bf \Pi}({\boldsymbol \Phi}_\text{uncon})$.}, the proposed methods in~\cite{SVDShrinkage} and~\cite{GDes}, as well as randomly, where the entries of ${\boldsymbol \Phi}$ are chosen from a zero-mean circularly-symmetric complex Gaussian distribution, termed EVD, Itr-SVD, GD, and Random, respectively. We show the simulation results in terms of the maximum mutual coherence $\mu_{\max}{({\bf \Psi})}$ defined in~\eqref{Mu} and the average mutual coherence $\mu_{\text{avg}}{({\bf \Psi})}$ defined as 
\begin{align}
	\mu_{\text{avg}}{({\bf \Psi})} = \frac{1}{N_\beta}\Big(\sum_{(k,j) \in \mathcal{S}_\beta}  \big|{\bf G}^{[k,j]}\big|\Big),
\end{align}
where $\mathcal{S}_\beta = \{ (k,j) \,:\, |{\bf G}^{[k,j]}\big| > \sqrt{\beta} \}$, $N_\beta$ is the number of elements in the set $\mathcal{S}_\beta$, and ${\bf G} = {\bf \Psi}^{\rm H} {\bf \Psi}$ is the normalized-diagonal Gram matrix. Table~\ref{MC_Tab} shows the obtained results for different values of $P$. Moreover, Fig.~\ref{conv1} shows the convergence behavior of the iterative methods for the scenarios with $P = 64$ and $P = 128$. For the GD method~\cite{GDes}, we use the step-size $\zeta = 5\times 10^{-4}/n$, while for the EGD method, we use $\zeta = 5\times 10^{-2}/n$, where $n$ is the iteration index.

\begin{figure}[tb]
	\centering
	\includegraphics[width=0.45\linewidth]{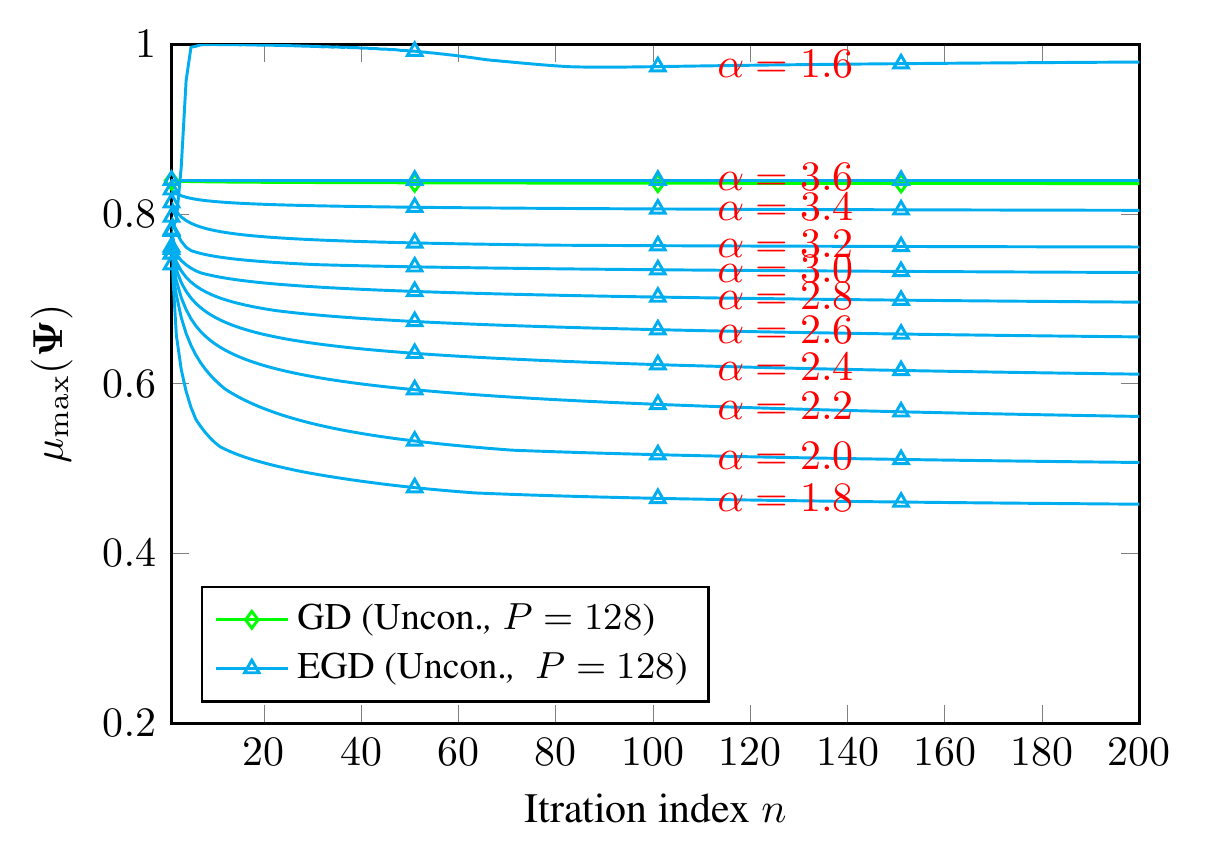}
	\caption{\small Coherence $\mu_{\max}{({\bf \Psi})}$ versus the iteration index.}
	\label{conv2}
\end{figure}

From Table~\ref{MC_Tab}, when $P = M = 64$, we can see that the SMCM and the Itr-SVD methods achieve similar performance, where the only difference is that SMCM has a faster convergence rate compared to Itr-SVD, as can be seen from Fig.~\ref{conv1}. However, as expected, when the ratio $P/M$ increases above 1, the SMCM performance decreases, since the na\"ive approach of calculating the mixing matrix ${\boldsymbol{\Phi}}$ from the designed sensing matrix ${\boldsymbol{\Psi}}$ incurs a performance loss. On the other hand, it can be seen that the proposed EGD method has the best performance in almost all of the considered scenarios. Here, we note that the introduced uncertainty measure $\alpha$ has a big impact on the EGD performance and the convergence rate, as can be seen from Fig.~\ref{conv2}. In general, for a sufficiently large $\alpha$, the EGD converges faster, but its performance degrades and approaches that of the GD. On the other hand, from Fig.~\ref{conv2}, we can also note that $\alpha$ should not be too small, since in this case most of the entries within the resulting error matrix $\tilde{{\bf E}}$ will be set to zero. From our simulation results in Table~\ref{MC_Tab}, we observe that $\alpha$ should be selected so that it is approximately equal to $P/M$.

\begin{table}[tb]
	\scriptsize
	\centering
	\caption{Coherence $\mu_{\max}{({\bf \Psi})}$ ($\mu_{\text{avg}}{({\bf \Psi})}$) versus $P$ ($N = 16 $ and $ M = 64$).}
	\label{MC_Tab}
	\begin{tabular*}{\textwidth}{@{}c@{\extracolsep{\fill}}ccccccc@{}}\toprule
		& $P$     & Random      & EVD         & Itr-SVD      & GD       & EGD        & SMCM   \\ \midrule
		\multicolumn{1}{l}{\multirow{3}{*}{${\boldsymbol \Phi}_\text{uncon}$}} & 64  & 0.64 (0.32) & 0.56 (0.30) & \textbf{0.24} (0.23) & 0.56 (0.31) & 0.26 (0.25) [$\alpha = 1.2$] & \textbf{0.24} (0.23) \\ 
		\multicolumn{1}{l}{}                               & 96   & 0.74 (0.33) & 0.74 (0.32) & 0.34 (0.25) & 0.67 (0.33) & \textbf{0.32} (0.30) [$\alpha = 1.4$] & 0.53 (0.28) \\ 
		\multicolumn{1}{l}{}                               & 128 & 0.85 (0.34) & 0.81 (0.33) & 0.50 (0.27) & 0.84 (0.34) & \textbf{0.44} (0.32) [$\alpha = 1.7$] & 0.73 (0.32) \\ \addlinespace
		\multirow{3}{*}{${\boldsymbol \Phi}_\text{con}$}                          & 64  & 0.64 (0.32) & 0.74 (0.32) & 0.51 (0.29) & 0.64 (0.31) & \textbf{0.31} (0.27) [$\alpha = 1.3$] & 0.57 (0.30) \\ 
		& 96  & 0.74 (0.33) & 0.75 (0.33) & 0.67 (0.31) & 0.68 (0.33) & \textbf{0.47} (0.30) [$\alpha = 1.5$] & 0.68 (0.33) \\
		& 128  & 0.85 (0.34) & 0.82 (0.34) & 0.79 (0.33) & 0.84 (0.34) & \textbf{0.72} (0.33) [$\alpha = 1.9$] & 0.80 (0.33) \\ \bottomrule
	\end{tabular*}
\end{table}


In this section, we have proposed the two mixing matrix design methods SMCM and EGD via mutual coherence minimization. For the unconstrained mixing matrix and $P = M$, we have shown that the original nonconvex problem can be relaxed and divided into $P$ convex subproblems, which are updated iteratively using an alternating optimization technique. However, SMCM incurs some performance loss for the constrained case and for $P > M$. To overcome this issue, we have proposed the EGD method, which enhances the classical GD-based method of~\cite{GDes} by introducing a shrinking operator on the error matrix. Using computer simulations, we have shown that the proposed SMCM and EGD methods have a faster convergence rate and a lower mutual coherence compared to the benchmark methods.

\section{Recovery Algorithms for Nonlinear Measurement Model}\label{sec:RecoveryAlgorithmsForNonlinearMeasurements}
This section is devoted recovery techniques that explicitly consider the specific structure of the measurements ${\bf z}$ themselves. More specifically, we consider the special case of magnitude only measurements. Hence, we use the information that measurements are nonnegative and we intend to uniquely recover the phase of the measurement signal along with the sparse representation vector.
\subsection{Phase Retrieval}
\label{subsec:phaseRetrieval}

In this subsection we consider the phase retrieval problem for a known dictionary ${\bf A}$, which aims to reconstruct an unknown complex-valued signal ${\bf x} \in \C^K$ from $M$ noise-corrupted magnitude-only measurements:
\begin{equation}
	\label{eq:phaseRetrieval}
	{\bf z} = \lvert{\bf A} {\bf x}\rvert + {\bf n},
\end{equation}
where ${\bf A}$ is a designed sensing matrix, ${\bf n} \in \C^M$ is an additive noise vector, and $\lvert\cdot\rvert$ is applied element-wise. The measurement model~\eqref{eq:phaseRetrieval} can be viewed as a special case of the system depicted in Fig.~\ref{HPPfig:linear_nonlinear_mixing_network} where $\boldsymbol{\Phi}$ is the identity and $\mathcal{T}\{ \cdot \} = \lvert\cdot\rvert$. Moreover, the original signal ${\bf x}$ is assumed to be sparse. Therefore, the recovery problem can be formulated as the following $\ell_1$ regularized nonlinear least-squares:
\begin{equation}
	\label{prob:phaseRetrieval}
	\underset{{\bf x} \in \C^K}{\min}\, h({\bf x}) =  \underbrace{\tfrac{1}{2} \left\lVert {\bf z} - \lvert{\bf A} {\bf x}\rvert \right\rVert_2^2}_{f({\bf x})} + \underbrace{\lambda \left\lVert {\bf x}\right\rVert_1}_{g({\bf x})}.
\end{equation}
It is a very challenging optimization problem due to the fact that $ g $ is nonsmooth and, more notably, $ f $ is nonsmooth and nonconvex. Besides, the original signal ${\bf x}$ can only be recovered up to a global phase ambiguity as ${\bf x} \cdot e^{j \phi}$ preserves both the magnitude measurements and the sparsity pattern.

We solve problem~\eqref{prob:phaseRetrieval} using the STELA algorithm in~\cite{yangParallelCoordinateDescent2019}, which is built on the majorization-minimization (MM) techniques in~\cite{qiuUndersampledSparsePhase2017} and the block successive convex approximation (BSCA) framework in~\cite{yangUnifiedSuccessivePseudoconvex2017,yangInexactBlockCoordinate2020}. The algorithm finds a stationary point of~\eqref{prob:phaseRetrieval} via a sequence of approximate problems that can be solved in parallel.
As~$f$ in the objective function of~\eqref{prob:phaseRetrieval} is nonconvex and nonsmooth, in each iteration we first construct a smooth upper bound function for $f$. 
Then, a descent direction of the upper bound function is obtained by solving a separable convex approximate problem, and a step-size along the descent direction is computed efficiently by exact line search. A decrease of the original objective function $h$ is ensured as its upper bound is decreased.
Let ${\bf x}^{(l)}$ be the current point in the $l$-th iteration. Specifically, the algorithm performs the following three steps in each iteration:
\begin{enumerate}
	\item \textbf{Smooth majorization.} The quadratic function $f$ in~\eqref{prob:phaseRetrieval} can be expanded as
	\begin{equation}
		f({\bf x}) = \tfrac{1}{2} \big(\lVert{\bf z}\rVert_2^2 + \lVert{\bf A}{\bf x}\rVert_2^2 \big) - {\bf z}^{\rm T} \lvert{\bf A} {\bf x}\rvert.
	\end{equation}
	Further, we note that for any $x \in \C$ and $\phi \in [0, 2 \pi)$
	\begin{equation}
		\label{eq:upbound}
		\left\lvert x\right\rvert = \lvert x \cdot e^{j \phi}\rvert \geq \Re \{ x \cdot e^{j \phi} \},
	\end{equation}
	and equality holds for $\phi = - \arg (x)$.
	Thus, defining 
	$ {\bf z}^{(l)} = {\bf z} \odot e^{j \arg ( {\bf A} {\bf x}^{(l)})}  $,
	where $e^{(\cdot)}$ and $\arg (\cdot)$ are applied element-wise and $ \odot $ denotes the Hadamard multiplication,
	we obtain the following smooth and convex upper bound for $f$ in the $l$-th iteration~\cite{qiuUndersampledSparsePhase2017}:
	\begin{align}
		\bar{f}^{(l)}({\bf x}) = \tfrac{1}{2} \big(\lVert{\bf z}\rVert_2^2 + \lVert{\bf A}{\bf x}\rVert_2^2 \big) - {\bf z}^{\rm T} \Re \big\{ {{\bf A} {\bf x}} \odot e^{-j \arg ( {\bf A} {\bf x}^{(l)})}  \big\} = \tfrac{1}{2} \lVert{\bf z}^{(t)} - {{\bf A} {\bf x}}\lVert_2^2,
	\end{align}
	which is tight at ${\bf x}^{(l)}$, i.e., $ \bar{f}^{(l)}({\bf x}^{(l)}) = f({\bf x}^{(l)}) $.
	Consequently, function
	$ \bar{h}^{(l)} ({\bf x}) = \bar{f}^{(t)} ({\bf x}) + g({\bf x}) $
	is also an upper bound of the objective function $h$ and tight at ${\bf x}^{(l)}$.
	
	\item \textbf{Descent direction computation.} Departing from the conventional MM algorithm, we minimize a separable convex approximation of $\bar{h}^{(l)}$, because $\bar{h}^{(l)}$ is computationally too expensive to minimize exactly for our present purpose. Based on the Jacobi algorithm~\cite{yangUnifiedSuccessivePseudoconvex2017}, 
	the convex approximate problem in the $t$-th iteration around point ${\bf x}^{(l)}$ is constructed as
	\begin{equation}
		\label{prob:phaseRetrievalApprox}
		\widetilde{{\bf x}}^{(l)} = \underset{{\bf x} \in \C^K}{\argmin}\, \sum_{k=1}^K \bar{f}^{(l)} \big( x_k,{\bf x}_{-k}^{(l)} \big) + g({\bf x}),
	\end{equation}
	where ${\bf x}_{-k}$ is a $(K-1)$-dimensional vector obtained by removing the $k$-th element $x_k$ from ${\bf x}$. 
	Problem~\eqref{prob:phaseRetrievalApprox} is decomposed into $K$ independent subproblems, which can be solved in parallel with suitable hardware~\cite{wangGPUacceleratedParallelOptimization2020}. Each subproblem is a Lagrangian form of single-variate LASSO, which admits a closed-form solution.
	According to~\cite[Prop. 1]{yangUnifiedSuccessivePseudoconvex2017}, the vector $\widetilde{{\bf x}}^{(l)} - {\bf x}^{(l)}$ represents a descent direction of $\bar{h}^{(l)}$. This motivates us to update ${\bf x}^{(l)}$ as follows
	\begin{equation}
		\label{eq:update}
		{\bf x}^{(l+1)} = {\bf x}^{(l)} + \gamma^{(l)} ( \widetilde{{\bf x}}^{(l)} - {\bf x}^{(l)}),
	\end{equation}
	where $\gamma^{(l)} \! \in \! [0,1]$ is the step-size. When $\widetilde{{\bf x}}^{(l)} = {\bf x}^{(l)}$, the algorithm has converged to a stationary point of $\bar{h}^{(l)}$, which is also stationary for the original problem~\eqref{prob:phaseRetrieval}~\cite[Thm. 1]{yangInexactBlockCoordinate2020}.
	\item \textbf{Step-size computation.} To efficiently find a proper step-size $\gamma^{(l)}$ for the update in~\eqref{eq:update}, we perform an exact line search on a differentiable upper bound of $\bar{h}^{(l)}$~\cite{yangUnifiedSuccessivePseudoconvex2017}.
	Thus, the computation of step-size $\gamma^t$ is formulated as
	\begin{equation}
		\label{prob:lineSearch}
		\gamma^{(l)} = \underset{0 \leq \gamma \leq 1}{\argmin} \; \bar{f}^{(l)} \big( {\bf x}^{(l)} + \gamma (\widetilde{{\bf x}}^{(l)} - {\bf x}^{(l)})\big) + g({\bf x}^{(l)}) + \gamma \big( g(\widetilde{{\bf x}}^{(l)}) - g({\bf x}^{(l)}) \big).
        \end{equation}
	The line search~\eqref{prob:lineSearch} corresponds to minimizing a convex quadratic function in the interval $[0,1]$, which can be solved in closed-form. Using the step-size $\gamma^{(l)}$ obtained by the line search~\eqref{prob:lineSearch} in the update~\eqref{eq:update}, a monotonic decrease of the original objective function $h$ in problem~\eqref{prob:phaseRetrieval} is ensured, cf.~\cite{yangParallelCoordinateDescent2019}.
\end{enumerate}

The mathematical expressions for the solutions of approximate problem~\eqref{prob:phaseRetrievalApprox} and line search~\eqref{prob:lineSearch} can be further found in~\cite{yangParallelCoordinateDescent2019}. Simulation results with Gaussian random sensing matrix ${\bf A}$ are also provided in~\cite{yangParallelCoordinateDescent2019}. The convergence analysis of the BSCA framework is presented in~\cite{yangInexactBlockCoordinate2020}. Besides, several other applications of the BSCA framework can be found in~\cite{yangEnergyEfficiencyMIMO2018,yangEnergyEfficiencyOptimization2019,yangOnlineParallelAlgorithm2016,yangParallelHybridSoftthresholding2018,liuBlockCoordinateDescent2019,liuParallelOptimizationApproach2019}. Furthermore, nonconvex regularization functions can be employed to resolve the defect that the $\ell_1$-regularization tends to produce biased estimates when the sparse signal has large coefficients~\cite{yangSuccessiveConvexApproximation2018}.

\subsection{Phase Retrieval with Dictionary Learning}
In the previous subsection, we considered the phase retrieval problem for signals that are sparse in the standard basis. However, in some cases, the signals that need to be recovered may only be sparse with respect to an unknown dictionary. Therefore, in this subsection we consider the phase retrieval with dictionary learning problem, which jointly learns a dictionary and sparse representations for reconstructing unknown signals~\cite{tillmannDOLPHInDictionaryLearning2016,qiuUndersampledSparsePhase2017,liuPARALLELALGORITHMPHASE2021}. 

As one possible application example, 
we consider a special case of the system depicted in Fig.~\ref{HPPfig:linear_nonlinear_mixing_network} with a known mixing matrix $\boldsymbol{\Phi}$ and $\mathcal{T} \{\cdot\} = \lvert\cdot\rvert$:
\begin{equation}
	{\bf z}(t) = \lvert \boldsymbol{\Phi} {\bf A} {\bf x}(t)\rvert + {\bf n}(t), \, t = 1, \ldots, D.
\end{equation}
Given $D$ time samples ${\bf Z} = [{\bf z}(1),\ldots,{\bf z}(D)]$, the objective is to jointly recover the unknown sensing matrix ${\bf A}$ and sparse transmitted signals ${\bf X} = [{\bf x}(1),\ldots,{\bf x}(D)]$.
The recovery problem is then formulated as the following phase retrieval with dictionary learning problem~\cite{liuPARALLELALGORITHMPHASE2021}:
\begin{align}
	\label{prob:phaseRetrievalDL}
	\underset{{\bf A} \in \mathcal{A}, {\bf X} \in \C^{K \times D}}{\min} \, h({\bf A},{\bf X}) =  \underbrace{\tfrac{1}{2} \left\lVert{\bf Z} - \lvert \boldsymbol{\Phi} {\bf A} {\bf X}\rvert\right\rVert_2^2}_{f({\bf A},{\bf X})} + \underbrace{\lambda \left\lVert{\bf X}\right\rVert_{1,1}}_{g({\bf X})}.
\end{align}
To avoid scaling ambiguities, we restrict ${\bf A}$ to be in the convex set $\mathcal{A} = \{{\bf A} \in \C^{M \times K} \,:\, \lVert{\bf a}_k\rVert_2 \leq 1, \forall k = 1, \ldots, K \} $. Also, $D > K$ is required to avoid trivial solutions.

Analogously, a stationary point of problem~\eqref{prob:phaseRetrievalDL} can be found by using the majorization technique in~\eqref{eq:upbound} and the BSCA framework. In addition to the procedure described in Section~\ref{subsec:phaseRetrieval}, we also partition the variables into two blocks, i.e., ${\bf A}$ and ${\bf X}$, and select a given number $k_B \in \{1,2\}$ of block variables to update in each iteration.
The block variables can be selected by cyclic or random update rules~\cite{yangInexactBlockCoordinate2020}.

Let $\big({\bf A}^{(l)}, {\bf X}^{(l)}\big)$ be the current point in the $l$-th iteration. We first consider the case where both block variables ${\bf A}$ and ${\bf X}$ are selected to update. Then, the three main steps that are performed in each iteration by the BSCA-based algorithm for problem~\eqref{prob:phaseRetrievalDL} are outlined as follows:
\begin{enumerate}
	\item \textbf{Smooth majorization.} 
	Exploiting the same majorization technique given in~\eqref{eq:upbound}, we construct a smooth upper bound for $f$ in~\eqref{prob:phaseRetrievalDL}.
	Defining 
	$ {\bf Z}^{(l)} = {\bf Z} \odot e^{j \arg ( \boldsymbol{\Phi} {\bf A} {\bf X}^{(l)})}  $,
	we can obtain the following smooth upper bound for $f$ in the $l$-th iteration:
	\begin{align}
		\label{eq:PRDL_upperBound}
		\bar{f}^{(l)}({\bf A},{\bf X}) 
		= \tfrac{1}{2} \lVert{\bf Z}^{(l)} - {\boldsymbol{\Phi} {\bf A} {\bf X}}\rVert_{\rm F}^2,
	\end{align}
	which is tight at $({\bf A}^{(l)},{\bf X}^{(l)})$.
	Similarly, we construct function
	$ \bar{h}^{(l)} ({\bf A},{\bf X}) = \bar{f}^{(l)} ({\bf A},{\bf X}) + g({\bf X}) $
	as an upper bound of the objective function $h$ which is tight at $\big({\bf A}^{(l)},{\bf X}^{(l)} \big)$. However, we remark that, unlike in Section~\ref{subsec:phaseRetrieval}, the upper bound function $\bar{f}^{(l)}$ in~\eqref{eq:PRDL_upperBound} is nonconvex due to the bilinear terms ${\bf A} {\bf X}$. Therefore, the convex approximation in the next step becomes necessary for efficiently finding a descent direction.
	
	\item \textbf{Descent direction computation.} Based on the Jacobi algorithm~\cite{yangUnifiedSuccessivePseudoconvex2017}, 
	the separable convex approximation for the minimization of $\bar{h}^{(l)}$ is constructed as
	\begin{equation}
		\label{prob:PRDLapprox}
		\big(\widetilde{{\bf A}}^{(l)}, \widetilde{{\bf X}}^{(l)} \big) \in \underset{{\bf A} \in \mathcal{A}, {\bf X}}{\argmin}\, \left\{
		\begin{array}{l}
			\sum_{m=1}^M \sum_{k=1}^K \bar{f}^{(l)} \big( x_{mk},{\bf A}^{(l)},{\bf X}_{-mk}^{(l)} \big) \\
			+\sum_{m=1}^M \bar{f}^{(l)} \big( {\bf a}_k, {\bf A}_{-k}^{(l)}, {\bf X}^{(l)} \big) + g({\bf X})
		\end{array} \right\},
	\end{equation}
	where ${\bf A}_{-k}$ is an $M \times (K-1)$ matrix obtained by removing the $k$-th column ${\bf a}_k$ from ${\bf A}$ and ${\bf X}_{-mk}$ denotes the collection of all entries of ${\bf X}$ except the $(m,k)$-th entry $x_{mk}$. 
	Problem~\eqref{prob:phaseRetrieval} can be decomposed into $K + (K \times D)$ independent subproblems. Each subproblem can be solved either in closed-form or by an efficient algorithm.
	Then, the difference $\big( \widetilde{{\bf A}}^{(l)} - {\bf A}^{(l)}, \widetilde{{\bf X}}^{(l)} - {\bf X}^{(l)} \big)$ represents a descent direction of $\bar{h}^{(l)}$ in the domain of problem~\eqref{prob:phaseRetrievalDL}. Defining $\Delta {\bf A} = \widetilde{{\bf A}}^{(l)} - {\bf A}^{(l)}$ and $\Delta {\bf X} = \widetilde{{\bf X}}^{(l)} - {\bf X}^{(l)}$, the following simultaneous update rule can be applied:
	\begin{equation}
		\label{eq:PRDLupdate}
		{\bf A}^{(l+1)} = {\bf A}^{(l)} + \gamma^{(l)} \Delta {\bf A} \quad \text{ and } \quad
		{\bf X}^{(l+1)} = {\bf X}^{(l)} + \gamma^{(l)} \Delta {\bf X},
	\end{equation}
	with a proper step-size $\gamma^{(l)} \in [0,1]$. When $\big(\widetilde{{\bf A}}^{(l)}, \widetilde{{\bf X}}^{(l)} \big) = \big({\bf A}^{(t)}, {\bf X}^{(l)} \big)$, the algorithm has converged to a stationary point of $\bar{h}^{(l)}$, which is also stationary for the original problem~\eqref{prob:phaseRetrievalDL}~\cite[Thm. 1]{yangInexactBlockCoordinate2020}.
	\item \textbf{Step-size computation.} We perform an exact line search on a differentiable upper bound of $\bar{h}^{(l)}$ to efficiently find a step-size $\gamma^{(l)}$ that ensures a monotonic decrease of the original objective function $h$ in~\eqref{prob:phaseRetrievalDL}.
	The computation of step-size $\gamma^{(l)}$ is then formulated as
	\begin{equation}
		\label{prob:PRDLlineSearch}
		\gamma^{(l)} = \underset{0 \leq \gamma \leq 1}{\argmin} \, \left\{
		\begin{array}{l}
			\bar{f}^{(l)} \big( {\bf A}^{(l)} + \gamma \Delta {\bf A}, {\bf X}^{(l)} + \gamma \Delta {\bf X} \big) \\ + g \big({\bf X}^{(l)} \big) + \gamma \left( g \big(\widetilde{{\bf X}}^{(l)} \big) - g \big({\bf X}^{(l)} \big) \right)
		\end{array} \right\}.
	\end{equation}
   	Problem~\eqref{prob:PRDLlineSearch} can be solved by rooting its derivative, a third-order polynomial, which admits a closed-form expression.
\end{enumerate}

In contrast to the above joint update case, if only one block variable is selected to update in the $l$-th iteration, 
then we solve the approximate problem~\eqref{prob:PRDLapprox} only with respect to the selected block variable,
which requires solving only the corresponding subproblems. 
Moreover, the update~\eqref{eq:PRDLupdate} is also performed only on the selected block variable, which is equivalent to setting the difference of the non-selected block variable to be all-zero. Further, when either of the matrices $\Delta {\bf A}$ and $\Delta {\bf X}$ is all-zero, the line search problem~\eqref{prob:PRDLlineSearch} reduces to a simple convex quadratic program.

Details of the BSCA-based algorithm for phase retrieval with dictionary learning and results from numerical experiments can further be found in~\cite{liuPARALLELALGORITHMPHASE2021}.

\section{Conclusions}\label{sec:ConclusionsAndOutlook}

Compressed sensing (CS) is a powerful technique for estimating sparse signals, which can be recovered, under mild conditions, from far fewer samples than otherwise indicated by the Nyquist-Shannon sampling theorem. Moreover, it was observed that incorporating side constraints not only improves the recovery guarantees but also reduces the required number of samples. This chapter builds on this important observation by addressing sparse signal reconstruction under various types of structural side constraints, including integrality, constant modulus, row and rank sparsity, and strict non-circularity constraints. Moreover, this chapter addresses the measurement system design for linear and nonlinear measurements of sparse signals. For the linear measurement systems, two mixing matrix design methods based on mutual coherence minimization are proposed, where constant modulus constraints are imposed element-wise to satisfy the mixing matrix hardware that involves cost-efficient analog phase shifters. For nonlinear measurement systems, parallel optimization design algorithms are proposed to efficiently compute the stationary points in the sparse phase retrieval problem with and without dictionary learning.

\bibliographystyle{spmpsci} 
\bibliography{bibfileHPP} 


\end{document}